\documentclass{article}
\usepackage[utf8]{inputenc}
\usepackage{amsthm,amsmath,amssymb,graphicx,gensymb}
\usepackage{natbib}
\RequirePackage[colorlinks,citecolor=blue,urlcolor=blue]{hyperref}
\usepackage[dvipsnames]{xcolor}
\usepackage{color}
\usepackage{bm}
\usepackage[margin=2cm]{geometry}
 
\usepackage{xr-hyper}
\makeatletter
\newcommand*{\addFileDependency}[1]{
  \typeout{(#1)}
  \@addtofilelist{#1}
  \IfFileExists{#1}{}{\typeout{No file #1.}}
}
\makeatother

\title{High-dimensional modeling of spatial and spatio-temporal conditional extremes using INLA and \edit{Gaussian Markov random fields}}
\author{E.\ S.\ Simpson\footnote{Corresponding author. Email address: emma.simpson@ucl.ac.uk} $^1$, T.\ Opitz$^2$, J.\ L.\ Wadsworth$^3$\\ 
\normalsize{$^1$Department of Statistical Science, University College London, Gower Street, London, WC1E 6BT, U.K.}\\
\normalsize{$^2$Biostatistics and Spatial Processes, INRAE, Avignon, France.}\\
\normalsize{$^3$Department of Mathematics and Statistics, Lancaster University, LA1 4YF, U.K.}} 
\date{\today}

\def\edit#1{{\textcolor{black}{#1}}}

\newcommand{\E}{\mathrm{E}}

\begin{document}

\maketitle
\begin{abstract}
The conditional extremes framework allows for event-based stochastic modeling of dependent extremes, and has recently been extended to spatial and spatio-temporal settings. After standardizing the marginal distributions and applying an appropriate linear normalization, certain non-stationary Gaussian processes can be used as asymptotically-motivated models for the process conditioned on threshold exceedances at a fixed reference location and time. \edit{In this work, we adapt existing conditional extremes models to allow for the handling of large spatial datasets. This involves specifying the model for spatial observations at $d$ locations in terms of a latent $m\ll d$ dimensional Gaussian model, whose structure is specified by a Gaussian Markov random field. We perform Bayesian inference for such models for datasets containing thousands of observation locations using the integrated nested Laplace approximation, or INLA.} We explain how constraints on the spatial and spatio-temporal Gaussian processes, arising from the conditioning mechanism, can be implemented through the latent variable approach without losing the computationally convenient Markov property. We discuss tools for the comparison of models via their posterior distributions, and illustrate the flexibility of the approach with gridded Red Sea surface temperature data at over $6,000$ observed locations. Posterior sampling is exploited to study the probability distribution of cluster functionals of spatial and spatio-temporal extreme episodes.
\end{abstract}


\subsubsection*{Acknowledgements}
This publication is based upon work supported by the King Abdullah University of Science and Technology (KAUST) Office of Sponsored Research (OSR) under Award No.\ OSR-2017-CRG6-3434.02. The sea surface temperature data studied in this paper were provided by GHRSST, Met Office and CMEMS.


\newpage
\section{Introduction}\label{sec:intro}

\subsection{Statistical modeling of spatial extremes}
The availability of increasingly detailed spatial and spatio-temporal datasets has motivated a recent surge in methodological developments to model such data. In this work, we are concerned with modeling extreme values of spatial or spatio-temporal processes, which we denote by $\{Y(s): s \in \mathcal{S} \subseteq \mathbb{R}^2\}$ and $\{Y(s,t): (s,t) \in \mathcal{S}\times\mathcal{T} \subseteq \mathbb{R}^2\times\mathbb{R}_+\}$. The goal of modeling spatio-temporal extremes is often to enable extrapolation from observed extreme values to future, more intense episodes, and consequently requires careful selection of models suited to this delicate task.

Early work on spatial extremes focused almost exclusively on max-stable processes \citep{Smith.1990,Coles.1993,Schlather.2002,Padoan.al.2010,Davison.Gholamrezaee.2012}. These are the limiting objects that arise through the operation of taking pointwise maxima of $n$ weakly dependent and identically distributed copies of a spatial process. However, this is a poor strategy when data exhibit a property known as \emph{asymptotic independence}, which means that the limiting process of maxima consists of everywhere-independent random variables. Moreover, even when the process is \emph{asymptotically dependent}, meaning that the limit has spatial coherence, the fact that the resulting process is formed from many underlying original events \citep{Dombry2018} can hinder both interpretability and inference. More recently, analogues of max-stable processes suited to event-level data have been developed \citep{Ferreira.deHaan.2014,Dombry.Ribatet.2015,Thibaud.Opitz.2016,deFondeville.Davison.2020}, but use of these generalized Pareto or $r$-Pareto processes also requires strong assumptions on the extremal dependence structure.

Broadly, spatial process data can be split according to whether they exhibit asymptotic independence or asymptotic dependence. As mentioned, these can be characterized by whether the data display independence or dependence in the limiting distribution of pointwise maxima, but when considering threshold exceedances other definitions are more useful. Consider two spatial locations $s, s+h \in \mathcal{S}$. For $Y(s) \sim F_s$, define the tail correlation function \citep{Strokorb2015} as
\begin{align}
\chi(s,s+h) = \lim_{q \to 1} \Pr\{F_{s+h}(Y(s+h))>q |F_{s}(Y(s))>q\}. \label{eq:chi}
\end{align}
If $\chi(s,s+h) = 0$ for all $h \neq 0$ then $Y$ is asymptotically independent, or asymptotically dependent where limit~\eqref{eq:chi} is positive for all $h$. Intermediate scenarios of asymptotic dependence up to a certain distance are also possible. Asymptotic dependence is a minimum requirement for use of max-stable or Pareto process models, but in practice more rigid assumptions are imposed as these models do not allow for any weakening of dependence with the level of the event --- a feature common in most environmental datasets.

Recent work on modeling of spatial extremes has focused on the twin challenges of incorporating flexible extremal dependence structures, and developing models and inference techniques that allow for large numbers of observation locations. \citet{Huser.al.2017,Huser.al.2020} suggest Gaussian scale mixture models, for which both types of extremal dependence can be captured depending on the distribution of the scaling variable. The model of \citet{Huser.Wadsworth.2018} was the first to offer a smooth transition between dependence classes, meaning it is not necessary to make a choice before fitting the model. However, owing to complicated likelihoods, each of these models is limited in practice to datasets with tens of observation locations. Modifications in \citet{Zhang.al.2019} suggest that hundreds of sites might be possible, but further scalability looks elusive for now.

\cite{Wadsworth.Tawn.2019} proposed an alternative approach based on a spatial adaptation of the multivariate conditional extreme value model \citep{Heffernan.Tawn.2004, Heffernan.Resnick.2007}, which has been further extended to the space-time case by \citet{Simpson.Wadsworth.2020}. Both types of extremal dependence can be handled and the likelihoods involved are much simpler. However, application thus far has still been limited to hundreds of observation locations. In this work, we seek to exploit the power of Gaussian Markov random fields and the integrated nested Laplace approximation (INLA) in this context in order to permit \edit{substantially higher} dimensional inference and prediction, and to achieve more flexible modeling by replacing parametric structures with semi-parametric extensions. We note that, in the restricted case of Pareto processes, \citet{deFondeville.Davison.2018} perform inference for a 3600-dimensional problem via a gradient-score algorithm. In this work, we handle inference for problems of comparable dimension, using the more flexible conditional extremes models with likelihood-based inference. \cite{Opitz.al.2018} have previously used INLA in an extreme value analysis context, focusing on regression modeling of threshold exceedances, but this is the first time it has been utilized in the conditional extremes framework.

\edit{In the remainder of the introduction, we provide background on the conditional extremes approach, and introduce briefly the idea of the INLA methodology. In Section~\ref{sec:modelFormulations}, we detail modifications to the conditional modeling approach that allow for inference through a latent variable framework using INLA; it is these tools that enable inference to become feasible at thousands of observation locations.}

\subsection{\edit{Conditional extremes models}}\label{subsec:CSEintro}
The aforementioned conditional extremes approaches, originating with \cite{Heffernan.Tawn.2004} in the multivariate case, involve the construction of models by conditioning on exceedances of a high threshold in a single variable. The spatial setting studied by \cite{Wadsworth.Tawn.2019}, and subsequent spatio-temporal extension of \cite{Simpson.Wadsworth.2020}, require conditioning on threshold exceedances at a single spatial or spatio-temporal location, with additional structure being introduced by exploiting the proximity of the other locations to this conditioning site.

In the spatial setting, denote by $\{X(s): s \in \mathcal{S}\}$ a stationary and isotropic process with marginal distributions possessing exponential-type upper tails, i.e., $\Pr\{X(s)>x\}\sim \edit{c}e^{-x}$ as $x\rightarrow\infty$, $\edit{c>0}$. This is achieved in practice via a marginal transformation, explained further in Section~\ref{subsec:redsea}. Let $s_0$ denote the conditioning site. We assume that $\{X(s)\}$ possesses a joint density, so that conditioning on the events $\{X(s_0)>u\}$ or $\{X(s_0)=u\}$ as $u \to \infty$ leads to the same limiting process \citep{Wadsworth.Tawn.2019}; see also \citet{Drees.Janssen.2017} for further discussion on the conditioning event in a multivariate setting. We comment on handling anisotropy in Section~\ref{sec:discussion}.

For a finite set of locations $s_1,\dots,s_d$, \cite{Wadsworth.Tawn.2019} assume that there exist normalizing functions $a_{s-s_0}(\cdot)$ and $b_{s-s_0}(\cdot)$ such that as $u\rightarrow\infty$,
\begin{align}
\Pr\left(\left[\frac{X(s_i)-a_{s_i-s_0}\left\{X(s_0)\right\}}{b_{s_i-s_0}\left\{X(s_0)\right\}}\right]_{i=1,\dots,d} \leq \bm{z}\bigg{\vert}~X(s_0)=u\right) ~\rightarrow \Pr\left(\{Z^0(s_i)\}_{i=1,\dots,d} \leq \bm{z}\right),
\label{eqn:CSEassumption}
\end{align}
where $\bm{z} = (z_1,\ldots, z_d)$, \edit{and the vector $\{Z^0(s_1),\ldots,Z^0(s_d)\}$ represents a finite-dimensional distribution of some stochastic process $\{Z^0(s)\}$, referred to as the residual process}. Several theoretical examples are provided therein to illustrate this assumption. The first of the normalizing functions is constrained to take values $a_0(x)=x$ and $a_{s-s_0}(x)\in[0,x]$, and is usually  non-increasing as the distance between $s$ and $s_0$ increases: the residual process therefore satisfies $Z^0(s_0)=0$. Furthermore, under assumption~\eqref{eqn:CSEassumption}, the excess of the conditioning variable $X(s_0)-u\mid X(s_0)>u$ is exponentially distributed, and independent of the residual process.

Assumption~\eqref{eqn:CSEassumption} is exploited for modeling by assuming that it holds approximately above a high threshold $u$. In particular, we can assume that
\begin{align}
\{X(s):s\in\mathcal{S}\} \mid \left[X(s_0)=x\right] = a_{s-s_0}(x) + b_{s-s_0}(x)\{Z^0(s):s\in\mathcal{S}\}, \qquad x>u.
\label{eqn:modelingAssumption}
\end{align}
Suitable choices for $a_{s-s_0}(\cdot),b_{s-s_0}(\cdot)$ and $\{Z^0(s)\}$ lead to models with different characteristics. \cite{Wadsworth.Tawn.2019} propose a theoretically-motivated parametric form for the normalizing function $a_{s-s_0}(\cdot)$, as well as three different parametric models for $b_{s-s_0}(\cdot)$ that are able to capture different tail dependence features. They propose constructing the residual process by first considering some stationary Gaussian process $\{Z(s)\}$, and either subtracting $Z(s_0)$ or conditioning on $Z(s_0)=0$ to ensure the condition $Z^0(s_0)=0$ on $\{Z^0(s)\}$ is satisfied. \edit{Marginal transformations of $\{Z^0(s)\}$ are considered therein in order to increase the flexibility of the models.} 

\edit{We note that assumptions~\eqref{eqn:CSEassumption} and \eqref{eqn:modelingAssumption} depend on the choice of $s_0$. In some applications, there may be a location of particular interest that would make a natural candidate for $s_0$, but for other scenarios the choice is not evident. However, under the assumption that $\{X(s)\}$ possesses a stationary dependence structure, in the sense that the joint distributions are invariant to translation, the form of the normalization functions $a_{s-s_0},b_{s-s_0}$ and the form of the residual process $\{Z^0(s)\}$ do not in fact depend on $s_0$, so that inference made using one conditioning location is applicable at any location. We discuss this issue further in Section~\ref{sec:sensitivity}.}

\edit{The approach to inference taken by \cite{Wadsworth.Tawn.2019} involves a composite likelihood. This allows different locations to play the role of the conditioning site, and combines information across each of these.} \edit{Inference under this ``vanilla'' version of the model, with $\{Z^0(s)\}$ constructed from a Gaussian process and parametric forms for the normalizing functions, can currently be performed for hundreds of observation locations.} However, scalability to thousands of locations is impeded by the $O(d^3)$-complexity of matrix inversion in the Gaussian process part of the likelihood and the fact that, in contrast to other areas of spatial statistics, we have $n$ replicates of the process to be used for inference. 

\subsection{\edit{INLA and the latent variable approach}}

\edit{In order to facilitate higher dimensional inference, we represent our model for observations at $d$ locations in terms of an $m$-dimensional latent Gaussian model, where $m \ll d$. This has the effect of creating a model that is amenable to use of the INLA framework, which allows for fast and accurate inference on the Bayesian posterior distribution of a parameter vector of interest, $\bm{\theta}$. It is particularly computationally convenient when the $m$-dimensional latent Gaussian component is endowed with a Gaussian Markov covariance structure, so that the precision matrix is sparse.}
	
\edit{The general form of the likelihood for models amenable to inference via INLA is 
	\begin{align}
	\pi(\bm{v}|\bm{\eta},\bm{\theta}) = \prod_{i=1}^d \pi(v_i|\eta_i,\bm{\theta}), \label{eq:inlamodel}
	\end{align}
where the observations $\bm{v} =(v_1,\ldots,v_d)^\top \in \mathbb{R}^d$, but the vector $\bm{\eta} = (\eta_1,\ldots,\eta_d)$ is a linear function of the $m$-dimensional latent Gaussian process. This specification of the distribution of $\bm{\eta}$, via a sparse precision matrix, allows for inference on the posterior distribution of interest, $\pi(\bm{\theta}|\bm{v})$, through a Laplace approximation to the necessary integrals. Note that the vector $\bm{\theta}$ also includes parameters of the latent Gaussian model, and is usually termed the \emph{hyperparameter} vector.}
	
\edit{A benefit of this Bayesian approach to statistical modeling with latent variables, over alternatives like the EM algorithm or Laplace approximations applied in a frequentist setting, is that parameters, predictions and uncertainties can be estimated simultaneously, and prior distributions can be used to incorporate expert knowledge and control the complexity of the model or its components with respect to simpler baselines. Moreover, the availability of the \texttt{R} software package \texttt{R-INLA} \citep{Rue.al.2017} facilitates the implementation, reusability and adaptation of our models and code, making them suitable for use with datasets other than the one considered in this paper. One of the main challenges we face is reconciling the form of the conditional extremes models with the formulation in equation~\eqref{eq:inlamodel} that is allowed under this framework. We outline our general strategy in Section~\ref{sec:modelFormulations}, but defer more detailed computational and implementation details to Section~\ref{sec:computation}. Our motivation for this is to provide readers with a general understanding of the methodology, unimpeded by extensive technical details. However, we note that implementation is a substantial part of the task and therefore Section~\ref{sec:computation} provides interested parties with the necessary particulars.}
	

\subsection{Overview of paper}
The remainder of the paper is structured as follows. \edit{In Section~\ref{sec:modelFormulations}, we provide a discussion on flexible forms for the conditional spatial extremes model that are possible under the latent variable framework. We discuss details of our inferential approach in Section~\ref{sec:spatialInference}, then apply this to a dataset of Red Sea surface temperatures in Section~\ref{sec:spatialApplication}, considering a range of diagnostics to aid model selection and the assessment of model fit. A spatio-temporal extension is presented in Section~\ref{sec:spacetimeInference}. Section~\ref{sec:computation} is aimed at those readers interested in the specifics of implementation, and includes more detail on INLA, the construction of Gauss-Markov random fields with approximate Mat\'ern covariance, and implementation of our models in \texttt{R-INLA}.} Section~\ref{sec:discussion} concludes with a discussion. Supplementary Material contains code for implementing the models we develop, and is available at \url{https://github.com/essimpson/INLA-conditional-extremes}.

\section{\edit{The latent variable approach and model formulations}}\label{sec:modelFormulations}
\subsection{Overview}
\edit{In this section, we begin by outlining details of the latent variable approach. We then build on the conditional extremes modeling assumption given in~\eqref{eqn:modelingAssumption} to allow for higher-dimensional inference under this latent variable framework. We discuss specific variants of the conditional extremes model that are possible in this case, summarizing the options in Section~\ref{subsec:modelSummary}.} 

\subsection{Generalities on the latent variable approach}\label{sec:generalities}
Here, we provide some general details on the latent variable approach for spatial modeling, denoting the observed data generically by $\bm V = (V_1,\ldots,V_d)^\top$, which in our context will correspond to observations at $d$ spatial locations. When modeling spatial extreme values, it is always necessary to have replications of the spatial process in question in order to distinguish between marginal distributions and dependence structures, and to define extreme events. We comment further on the handling of temporal replication in Sections~\ref{subsec:spacetime} and~\ref{sec:implementation}; we will later also explicitly model temporal, as well as spatial, dependence.

In hierarchical modeling with latent Gaussian processes, we define a latent, unobserved Gaussian process $\bm W=(W_1,\ldots,W_m)^\top$, with $m$ denoting the number of \edit{`locations'} of the latent process. \edit{These could encompass the spatial locations used to discretize the spatial domain, or the knots used in spline functions, for example.} We assume conditional independence of the observations $\bm V$ with respect to $\bm W$, and use the so-called \emph{observation matrix} $A\in\mathbb{R}^{d\times m}$ to define a linear predictor 
$$
\bm\eta = \bm\eta (\bm W) = A \bm W 
$$
that linearly combines the  latent variables in $\bm W$ into components $\eta_i$ associated with $V_i$, $i=1,\ldots,d$. The components $\eta_i$ represent a parameter of the probability distribution of $V_i$. The matrix $A$  is deterministic and is fixed before estimating the model. For instance, $A$ handles the piecewise linear spatial interpolation from the \edit{$m$ locations represented by the latent Gaussian vector $\bm W$ towards the $d$ observed sites; for this, $\bm W$ may contain the values of a spatial field} at locations $\tilde{s}_1,\ldots,\tilde{s}_m$, and $A$ has $i$-th line $A_i=(0,\ldots,0,1,0,\ldots,0)$ if the observation location $s_i$ of $V_i$ coincides with one of the locations $\tilde{s}_{j_0}$, where the $1$-entry is at the $j_0$th position. Otherwise, several entries of $A_i$ could have non-zero weight to implement interpolation between the $\tilde{s}_{j}$-locations. The distribution of $\bm\eta$ is also multivariate Gaussian due to the linear transformation.  The univariate probability distribution of $V_i$, often referred to as the \emph{likelihood model}, can be Gaussian or non-Gaussian and is parametrized by the linear predictor $\eta_i$, and potentially by other hyperparameters. The vector of hyperparameters (i.e., parameters that are not components of one of the Gaussian vectors $\bm W$ and $\bm \eta$), such as those related to variance, spatial dependence range, or smoothness of a spline curve, is denoted by $\bm \theta$. Letting $\pi(\cdot)$ denote a generic probability distribution, the hierarchical model is structured as follows:
\begin{align*}
\bm \theta &\sim \pi(\cdot) & \text{hyperparameters,}\\
\bm W\mid \bm \theta & \sim \mathcal{N}_m\left(\bm 0,  Q(\bm
\theta)^{-1}\right) &\text{latent Gaussian components,} \\
V_i\mid \bm W, \bm \theta &\sim   \pi(\cdot \mid \eta_i, \bm
\theta), ~~\mbox{independent}  &\text{likelihood\ of\ observations.}
\end{align*}
The matrix $Q(\bm\theta)$ denotes the precision matrix of the latent Gaussian vector $\bm W$, whose variance-covariance structure may depend on some of the hyperparameters in $\bm \theta$ that we seek to estimate. 
In the case of  observations $V_i$ having a Gaussian distribution\edit{, we can set the Gaussian mean as the linear predictor $\eta_i$. Then,} the conditional variance $\sigma^2$ of $V_i$ given $\eta_i$ is a hyperparameter, and we define
\begin{align}
V_i \mid \eta_i, \sigma^2  \sim \mathcal{N}(\eta_i, \sigma^2), \quad i=1,\ldots,d. 
\label{eqn:latentV}
\end{align}
\edit{Under the latent variable framework, if $d>m$ as in our setting, we always need a small positive variance $\sigma^2>0$ in~\eqref{eqn:latentV}, since there is no observation matrix $A$ that would allow for an exact solution to the equation $\bm V = A\bm W$ with given $\bm V$. The presence of this independent component with positive variance  is therefore a necessity for the latent variable approach. In some modeling contexts it can be interpreted as a useful model feature, e.g., to capture measurement errors. We discuss the consequences of this $\sigma^2$ parameter on our conditional model in Section~\ref{sec:spatialModels}. With the Gaussian likelihood \eqref{eqn:latentV}, the multivariate distribution of $\bm V$ given  $\sigma^2$ is still Gaussian, and in this case the Laplace approximation to the posterior $\pi(\bm{\theta}|\bm{v})$ with the observation $\bm{v}$ of $\bm V$  is exact and therefore does not induce any approximation biases.}

A major benefit of the construction with latent variables is that the dimension of the latent vector $\bm W$ is not directly determined by the number of observations $d$. The computational complexity and stability of matrix operations (e.g., determinants, matrix products, solution of linear systems) arising in the likelihood calculations for the above Bayesian hierarchical  model is therefore mainly determined by the tractability of the precision matrix $Q(\bm\theta)$, whose dimension can be controlled independently from the number of observations. Such matrix operations can be implemented very efficiently if precision matrices are sparse \citep{Rue.Held.2005}. If data are replicated with dependence between replications, such as spatial data observed at regular time steps in spatio-temporal modeling, the sparsity property can be preserved in the precision matrix of the latent space-time process $\bm W$. In this work, we will make assumptions related to the separability of space and time, which allows us to generate sparse space-time precision matrices by combining sparse precision matrices of a purely spatial and a purely temporal process \edit{using the Kronecker product of the two matrices}; see Sections~\ref{subsec:spacetime} and~\ref{sec:implementation} for further details.

\subsection{\edit{The latent variable approach for conditional extremes model inference}}\label{sec:spatialModels}
\edit{Here, we explain how the latent variable approach can be used within the conditional extremes framework to reduce the dimension of the residual process $\{Z^0(s)\}$ for inferential purposes. We begin by presenting the conditional extremes model with parametric forms for the normalizing functions $a_{s-s_0}$ and $b_{s-s_0}$, following the approach of \cite{Wadsworth.Tawn.2019}. In Section~\ref{sec:proposed.a}, we propose a more flexible semi-parametric form for $a_{s-s_0}$ that further exploits the latent variable framework.}

\edit{Consider the conditional extremes model presented in~\eqref{eqn:modelingAssumption}. Fixing $a_{s-s_0}(x)=x$ and $b_{s-s_0}(x)=1$ enforces asymptotic dependence, but setting $a_{s-s_0}(x) = x\alpha(s-s_0)$ and allowing the form of $\alpha(s-s_0)$ to depend on the distance from the conditioning location is the key aspect that enables modeling of asymptotic independence as well. To capture asymptotic independence, \cite{Wadsworth.Tawn.2019} propose a parametric form for $\alpha(\cdot)$, defining
\begin{align}
    \alpha(s-s_0) = \exp\left\{-\left(\|s-s_0\|/\lambda\right)^\kappa\right\},\qquad \lambda>0, \qquad 0\leq\kappa\leq 2.
\label{eqn:parametric.a}
\end{align}
The resulting function $a_{s-s_0}$ satisfies the constraint that $a_0(x)=x$ and has $a_{s-s_0}(x)$ decreasing as the distance to $s_0$ increases. \cite{Wadsworth.Tawn.2019} propose three different parametric forms for the normalizing function $b_{s-s_0}$, each with different modeling aims. We focus on the option of $b_{s-s_0}(x)=x^\beta$, with $\beta\in[0,1)$, throughout the rest of the paper, including the simple special case where we fix $\beta=0$.}
 
\edit{The existing conditional extremes approach must be simplified and slightly modified to allow for inference using the latent variable framework. First, in Section~\ref{subsec:CSEintro}, we discussed that \cite{Wadsworth.Tawn.2019} consider marginal transformations of the residual process to increase the flexibility of their approach. A special case of this is to simply restrict $\{Z^0(s)\}$ to have Gaussian margins. This is the approach we take in order to adopt the framework described in Section~\ref{sec:generalities} and facilitate computation. Moreover, as highlighted in equation~\eqref{eqn:latentV}, the use of a latent process of dimension $m<d$ requires the introduction of a variance term $\sigma^2>0$ common to each of the observation locations. Taking this into consideration, our conditional extremes model now has the form
\begin{align}
X(s_i)\mid [X(s_0)=x] = x\alpha(s_i-s_0) + x^\beta Z^0(s_i) + \epsilon_i, \qquad \epsilon_i \sim \mathcal{N}(0,\sigma^2)~~ \mbox{i.i.d.},
\label{eqn:latentConditionalForm}
\end{align}
for $i=1,\dots,d$, where we assume $\epsilon_0 = 0$ at $s_0$. A key point here is that the Gaussian noise does not represent a model feature to capture measurement error or add extra roughness to the process; it is simply included for computational feasibility.}  

\edit{The latent variable approach described in Section~\ref{sec:generalities} can be applied to the residual process in~\eqref{eqn:latentConditionalForm}, providing us with a ``low-rank'' representation of $\{Z^0(s)\}$. The constraint that $Z^0(s_0)=0$ can be enforced by manipulating the observation matrix $A$; further detail on this is provided in Section~\ref{subsec:residualConstraint}. In this case, assuming the parametric form~\eqref{eqn:parametric.a} for $\alpha(s-s_0)$ requires estimation of the parameters $(\lambda,\kappa)$, in addition to the parameter $\beta$ of the $b_{s-s_0}$ function. Under the latent variable framework, these are included as part of the hyperparameter vector $\bm \theta$. The dimension of $\bm{\theta}$ must remain moderate (say, at most $10$ to $20$ components), since INLA requires numerical integration with integrand functions defined on the hyperparameter space. In the implementation of INLA using \texttt{R-INLA}, estimation of the parameters of $(\lambda,\kappa,\beta)$ requires the use of specific user-defined (``\texttt{generic}'') models}, which we describe in Section~\ref{sec:implementation}. We emphasize that the use of \texttt{generic} \texttt{R-INLA} models allows for the implementation of other relevant parametric forms for the functions $a_{s-s_0}$ and $b_{s-s_0}$, if the above choices do not provide a satisfactory fit.

\edit{For the function $\alpha$, an alternative to parametric forms is to adopt a semi-parametric approach by constructing $x \alpha(\cdot)$ as an additive contribution to the linear predictor with multivariate Gaussian prior distribution. However, the function $b_{s-s_0}(x)$ must have a parametric form with a small number of parameters included in the hyperparameter vector $\bm\theta$, because in the INLA framework it is not possible to represent both $b_{s-s_0}(x)$ and $\{Z^0(s)\}$ as latent Gaussian components, given the restriction of using a linear transformation from $\bm W$ to $\bm \eta$. This is achieved by our choice to set $b_{s-s_0}(x)=x^\beta$.} \edit{Some frequentist estimation approaches for generalized additive models implement Laplace approximation techniques where semiparametric forms of the variance of the Gaussian likelihood are possible \citep{Wood2011}. However, this approach is currently not available within \texttt{R-INLA} and may come at the price of less stable estimation due to identifiability problems and less accurate Laplace approximations, and estimation can become particularly cumbersome with large datasets, such that we do not pursue it here.}

\subsection{\edit{Semi-parametric modeling of $a_{s-s_0}$}}\label{sec:proposed.a}
\edit{Continuing to adopt the form $a_{s-s_0}(x) = \alpha(s-s_0)x$, we now consider semi-parametric modeling of $\alpha$. In subsequent sections} we focus on this solution for its novelty, increased flexibility and computational convenience. \edit{Semi-parametric forms can be implemented by using a B-spline function for $\alpha(s-s_0)$, which forms an additive component of the linear predictor $\bm \eta$. This is computationally convenient since INLA can handle a large number of latent Gaussian variables in $\bm W$ when calculating accurate deterministic approximations to posterior distributions, via the Laplace approximation.  We constrain this function to have $\alpha(0)=1$, ensuring that $a_{0}(x)=x$.}

In extension to the models for conditional spatial and spatio-temporal extremes developed by \citet{Wadsworth.Tawn.2019} and \citet{Simpson.Wadsworth.2020}, we can further increase the flexibility of the conditional mean model by explicitly including a second spline function, denoted $\gamma(s-s_0)$ and with $\gamma(0)=0$, that is not multiplied by the value of the process at the conditioning site. \edit{To clarify, this implies that we have $a_{s-s_0}(x)=\alpha(s-s_0)x+\gamma(s-s_0)$, with $\gamma(s-s_0)$ also incorporated as a component of the linear predictor $\bm \eta$.} An example where such a deterministic component arises is given by the conditional extremes model corresponding to the Brown--Resnick type max-stable processes \citep{Kabluchko.Schlather.deHaan.2009} with log-Gaussian spectral function \citep[see Proposition 4 of][]{Dombry.al.2016}, which are widely used in statistical approaches based on the asymptotically dependent limit models mentioned in Section~\ref{sec:intro}; in this case, we obtain 
\[
X(s)\mid [X(s_0)=x] \stackrel{d}{=} x + Z(s)-Z(s_0)-\text{Var}(Z(s)-Z(s_0))/2,
\]
with a centered Gaussian process $\{Z(s)\}$. \edit{Therefore, by setting $\alpha(s-s_0)=1$}, in this model the $\gamma$-term corresponds to the semi-variogram $\text{Var}(Z(s)-Z(s_0))/2$. We note that for the Brown--Resnick process, $\gamma$ should indeed correspond to a valid semi-variogram, although we will not constrain it as such in our implementation \edit{to allow for greater flexibility. However, we underline that in the INLA framework there is no impediment to using parametric forms of $\gamma$ with parameters included in the hyperparameter vector $\bm{\theta}$.}

\subsection{Proposed models}\label{subsec:modelSummary}
\edit{To summarize, in the implementation of the conditional spatial extremes modeling assumption~\eqref{eqn:latentConditionalForm} using \texttt{R-INLA}, we propose to explore several options for the form of the model:} setting $\alpha(s-s_0)=1$ everywhere or using a spline function; whether or not to include the second spline term $\gamma(s-s_0)$; and whether or not to include the parameter $\beta$. Together, this means that all models can be written as special cases of the representation
\begin{align}
X(s_i)\mid [X(s_0)=x] = \alpha(s_i-s_0) x + \gamma(s_i-s_0) + x^\beta Z^0(s_i) + \epsilon_i, \qquad x>u, \qquad i=1,\ldots,d, \label{eq:modelgeneral}
\end{align}
where we suppose that $\{Z^0(s)\}$ has a Gaussian structure, further described in Sections~\ref{subsec:SPDE} and~\ref{subsec:residualConstraint}, and $\epsilon_i \sim \mathcal{N}(0,\sigma^2)$ i.i.d.. This opens up the framework of conditional Gaussian models and the potential for efficient inference via INLA, while closely following the conditional extremes formulation. In particular, the joint distribution of $\{X(s_i): i=1,\ldots, d\}$, not conditional on the value of $X(s_0)$, is non-Gaussian. 

\edit{Finally, we give an illustration, linking model~\eqref{eq:modelgeneral} to the general notation and principles outlined in Section~\ref{sec:generalities}.} Our observation vector $\bm{V}$ is the process $\{X(s)\}$ observed at $d$ locations: $\bm{X}=(X(s_1),\ldots,X(s_d))^\top$. The latent Gaussian component $\bm{W}$ consists of components for $\alpha,\gamma$ and $\{Z^0(s)\}$: $\bm{W} = (\bm{W}_\alpha^\top, \bm{W}_\gamma^\top, \bm{W}_Z^\top)^\top \in \mathbb{R}^{m_{\alpha}} \times \mathbb{R}^{m_\gamma} \times \mathbb{R}^{m_Z}$, with $m_\alpha+m_\gamma+m_Z = m$. The observation matrix $A \in \mathbb{R}^{d \times m}$ is the concatenation of matrices for each component: $A_\alpha \in \mathbb{R}^{d \times m_\alpha}$, $A_\gamma \in \mathbb{R}^{d \times m_\gamma}$, and $A_S \in \mathbb{R}^{d \times m_Z}$. We include the $x^\beta$-term into the process $\bm{W}_Z$ if we  want to estimate the parameter $\beta$, such that it does not appear in the fixed observation matrix $A_S$; if $\beta$ is fixed, we could instead include the $x^\beta$-term into $A_S$. 

\edit{We emphasize that the model is applied to replicates of the observed process $\bm{X}$ and that while the $\alpha$, $\gamma$ and $\beta$ components are fixed, the residual process and error term generally vary across replicates. All together for the $j$th replicate $\bm{X}_j$, we get
\[
\bm{X}_j|[X_j(s_0)=x_j] = (x_j A_\alpha,A_\gamma,A_{S})(\bm{W}_\alpha^\top, \bm{W}_\gamma^\top, \bm{W}_{Z,j}^\top)^\top + \bm{\epsilon}_j,
\]
with i.i.d.\ Gaussian components in $\bm{\epsilon}_j = (\epsilon_{1,j},\ldots,\epsilon_{d,j})^\top$, and where the $j$-subscripts highlight the components that vary with replicate.} \edit{To implement model~\eqref{eq:modelgeneral} in an efficient manner for a large number of observation locations, we need to carefully consider computations related to the residual process $\{Z^0(s)\}$; this is explained in detail in Section~\ref{subsec:residualConstraint}.}

\section{\edit{Inference for conditional spatial extremes}}\label{sec:spatialInference}
\subsection{\edit{Overview}}
\edit{ In Section~\ref{sec:spatialApplication}, we apply variants of model~\eqref{eq:modelgeneral} to the Red Sea surface temperature data, with the different model forms summarized in Table~\ref{tab:spatialModels}. In this section, we discuss certain considerations necessary to carry out inference and techniques to compare the candidate models. In Section~\ref{subsec:marginalTransform}, we begin with a discussion of the transformation to exponential-tailed marginal distributions that are required for conditional extremes modeling. We discuss construction of the observation matrix $A$ and choices of prior distributions for the hyperparameters in Section~\ref{sec:hyperprior}. In Sections~\ref{subsec:spatialModelSelection}~and~\ref{subsec:loocv}, we present two approaches for model selection and validation, both of which are conveniently implemented in the \texttt{R-INLA} package and therefore straightforward to apply in our setting.}

\subsection{\edit{Marginal transformation}}\label{subsec:marginalTransform}
To ensure the marginal distributions of the data have the required exponential upper tails, we suggest transforming to Laplace scale, as proposed by \cite{Keef.al.2013}. This is achieved using a semiparametric transformation. Let $Y$ denote the surface temperature observations at a single location. We assume these observations follow a generalized Pareto distribution above some high threshold $v$ to be selected, and use an empirical distribution function below $v$, denoted by $\tilde{F}(\cdot)$. That is, we assume the distribution function
\begin{align}
F(y)=
\begin{cases}
1 - \lambda_v\left\{1+\frac{\xi(y-v)}{\sigma_v}\right\}^{-1/\xi}_+, &y\geq v\\
\tilde{F}(y), &y<v,
\end{cases}
\label{eqn:marginalDist}
\end{align}
for $\lambda_v=1-F(v)$, $\sigma_v>0$ and $y_+=\max(y,0)$. Having fitted this model, we obtain standard Laplace margins via the transformation
\[
X=
\begin{cases}
\log\left\{2F(Y)\right\}, &F(Y)\leq1/2,\\
-\log\left[2\{1-F(Y)\}\right], &F(Y)>1/2.
\end{cases}
\]
\edit{This transformation should be applied separately for each spatial location, and we estimate the parameters of the generalized Pareto distributions using the \texttt{ismev} package in \texttt{R} \citep{ismev}.} It is possible to include covariate information in the marginal parameters and impose spatial smoothness on these, for instance by using flexible generalized additive models (see \citet{CastroCamilo2020} for details), but we do not take this approach. 

\subsection{\edit{Spatial discretization and prior distributions for hyperparameters}}\label{sec:hyperprior}

\edit{We now discuss the distribution of the latent processes $\bm{W}_\alpha, \bm{W}_\gamma$ and $\bm{W}_Z$, as defined in Section~\ref{subsec:modelSummary}. Gauss--Markov distributions for these components, with approximate Mat\'ern covariance, are achieved through the \edit{stochastic partial differential equation (SPDE)} approach of \citet{Lindgren.al.2011}. The locations of the components of the multivariate Gaussian vector $\bm{W}_Z$ defining the latent spatial process are placed at the nodes of a triangulation covering the study area. To generate this spatial discretization of the latent Gaussian process, and the observation matrix $A$ to link it to observations,} \edit{we use a mesh. An example of this will be discussed for our Red Sea application in Section~\ref{subsec:RedSeaMeshPriors} and demonstrated pictorially in Figure~\ref{fig:largeSpatialMesh}.} \edit{Full technical details of the construction of spatial and spatio-temporal precision matrices $Q$ for each component, and observation matrices $A$, are provided in Section~\ref{sec:computation}.}

\edit{For the one-dimensional spline functions we propose for modeling $\alpha(s-s_0)$ and $\gamma(s-s_0)$, we suggest choosing knots that are equally spaced and relate to the distance from the conditioning site, $s-s_0$. One knot should be placed at a distance of $s-s_0=0$ to allow us to enforce the constraints that $\alpha(0)=1$ and $\gamma(0)=0$. We use quadratic spline functions for both $\alpha(s-s_0)$ and $\gamma(s-s_0)$, which we have found to provide more flexibility than their linear counterparts. For the SPDE priors corresponding to these spline components, we suggest fixing the range and standard deviation, since we consider that estimating these parameters is not crucial. This avoids the very high computational cost that can arise when we estimate too many hyperparameters with \texttt{R-INLA}.}

For specifying the prior distributions of the hyperparameters (e.g., variances, spatial ranges, autocorrelation coefficients \edit{for the space-time extension}) we use the concept of penalized complexity (PC) priors \citep{Simpson.al.2017}, which has become the standard approach in the INLA framework.  PC priors control model complexity by shrinking model components towards a simpler baseline model, using a  constant rate penalty expressed through the Kullback-Leibler divergence of the more complex model with respect to the baseline. In practice, only the rate parameter has to be chosen by the modeler, and it can be determined indirectly---but in a unique and intuitive way---by setting a threshold value $r$ and a probability $p\in (0,1)$  such that $\mathrm{Pr}(\text{hyperparameter}>r)=p$, with $>$ replaced by $<$ in some cases, depending on the role of the parameter. For example, the standard baseline model for a variance parameter of a latent and centered Gaussian prior component $\bm W$ contributing to the linear predictor $\bm \eta$  is a variance of $0$, which corresponds to the absence of this component from the model, and the PC prior of the standard deviation corresponds to an exponential distribution. Analogously, we can fix the PC prior for the variance parameter $\sigma^2$ of the observation errors $\epsilon_i$ in \eqref{eq:modelgeneral}. \edit{If the data-generating process is smooth then $\sigma^2$ could instead be fixed to a very small value, but for reasons of stabilizing the estimation procedure, we prefer to estimate its value from the data.} For the Mat\'ern covariance function, PC priors are defined jointly for the range and the variance, with the baseline given by infinite range and variance $0$; in particular, the inverse of the range parameter has a PC prior given by an exponential distribution, see \citet{Fuglstad.al.2019} for details. As explained by \citet{Simpson.al.2017}, PC priors are not useful to ``regularize" models, i.e., to select a moderate number of model components among a very large number of possible model components. Rather, they are used to control the complexity of a moderate number of well-chosen components that always remain present in the posterior model, and they do not put any positive prior mass on the baseline model.

\subsection{Model selection using WAIC}\label{subsec:spatialModelSelection}
Conveniently, implementation in \texttt{R-INLA} allows for automatic estimation of certain information criteria that can be used for model selection. Two such criteria are the deviance information criterion (DIC), and the widely applicable or Watanabe-Akaike information criterion (WAIC) of \cite{Watanabe2013}. We favour the latter since it captures posterior uncertainty more fully than the DIC. This, and other, benefits of the WAIC over the DIC are discussed by \cite{Vehtari.etal.2017}, where an explanation of how to estimate the WAIC is also provided. Using our general notation for latent variable models, as in Section~\ref{sec:generalities}, suppose that the posterior distribution of the vector of model parameters $\bm{\tilde{\theta}}=(\bm \theta^\top, \bm W^\top)^\top$ is represented by a sample $\tilde{\bm\theta}^s$, $s=1,\dots,S$, with the corresponding sample variance operator given by $\mathbb{V}^S_{s=1}(\cdot)$. Given the observations $v_i$, $i=1,\ldots,d$, the WAIC is then estimated as
\[
\sum_{i=1}^d\log\left\{\frac{1}{S}\sum_{s=1}^S\pi(v_i\mid \tilde{\bm\theta}^s)\right\} - \sum_{i=1}^d \mathbb{V}_{s=1}^S\left\{\log \pi(v_i\mid \tilde{\bm\theta}^s)\right\},
\] 
with the first term providing an estimate of the log predictive density, and the second an estimate of the effective number of parameters. Within \texttt{R-INLA}, we do not generate a representative sample, but the sample means and variances  with respect to $\tilde{\bm\theta}^s$ in the above equation are estimated based on \texttt{R-INLA}'s internal representation of posterior distributions; see also the estimation technique for the DIC explained in \citet[][Section~6.4]{Rue.al.2009}. \edit{Smaller values of the WAIC indicate more successful model fits, and we will use this criterion to inform our choice of model for the Red Sea data in Section~\ref{subsec:spatialModels}.} 

\subsection{\edit{Cross validation procedures}}\label{subsec:loocv}

As mentioned previously, the main aim of fitting conditional extremes models is usually to extrapolate to extreme levels that have not been previously observed. However, the INLA framework also lends itself to the task of interpolation, e.g., making predictions for unobserved locations. Although interpolation is not our aim, here we discuss some procedures that allow for the assessment of models in this setting. For model selection, we can use cross-validated predictive measures, based on leave-one-out cross-validation (LOO-CV). These are relatively quick to estimate with INLA without the need to re-estimate the full model; see \citet[][Section~6.3]{Rue.al.2009}. Here, one possible summary measure is the \emph{conditional predictive ordinate} (CPO), corresponding to the predictive density of observation $v_i$ given all the other observations $\bm v_{-i}$, i.e., 
$$
\text{CPO}_i = \pi(v_i\mid \bm v_{-i}),
$$
for $i=1,\dots,d$. Log-transformed values of CPO define the log-scores often used in the context of prediction and forecasting \citep{Gneiting2007}. A model with higher CPO values usually indicates better predictions. We note that the CPO is not usually used for extreme value models \edit{where interpolation is often not considered the main goal.} \edit{It will not be particularly informative in our application} since the loss of information from holding out a single observation is negligible in the case of densely observed processes with very smooth surface. However, we include it as it may be useful for other applications where spatial and temporal interpolation are important, \edit{for example when data are observed at irregularly scattered meteorological stations,} and due to the simplicity of its calculation in \texttt{R-INLA}.

One can also consider the \emph{probability integral transform} (PIT) corresponding to the distribution function of the predictive density, evaluated at the observed value $v_i$, i.e., 
$$
\text{PIT}_i = \int_{-\infty}^{v_i}\pi(v\mid \bm v_{-i})\mathrm{d}v.
$$
If the predictive model for single hold-out observations appropriately captures the variability in the observed data, then the PIT values will be close to uniform. \edit{If in a histogram of PIT values the mass concentrates strongly towards the boundaries, then predictive credible intervals (CIs) will be too narrow; by contrast, if mass concentrates in the middle of the histogram, then predictive CIs will be too large.} \edit{We refer the reader to \citet{Czado2009predictive} for more background on PITs. We discuss such cross validation procedures in Section~\ref{subsec:spatialModels}.}


\section{Application to modeling Red Sea surface temperature extremes}\label{sec:spatialApplication}

\subsection{Overview}
\edit{In this section, we propose specific model structures for the general model \eqref{eq:modelgeneral} and illustrate an application of our approach using a dataset of Red Sea surface temperatures, the spatio-temporal extremes of which have also been studied by \cite{Hazra.Huser.2020} and \cite{Simpson.Wadsworth.2020}, for instance. We focus on the purely spatial case here, and consider further spatio-temporal modeling extensions for this dataset in Section~\ref{sec:spacetimeInference}. We use the methods discussed in Sections~\ref{subsec:spatialModelSelection}~and~\ref{subsec:loocv} to assess the relative suitability of the proposed models for the Red Sea data.} For the best-fitting model, we present additional results, and conclude with a discussion of consequences of using a single, fixed conditioning site. Throughout this section, the threshold $u$ in conditional model~\eqref{eq:modelgeneral} is taken to be the 0.95 quantile of the transformed data, following \cite{Simpson.Wadsworth.2020}. For the sake of a brevity, we do not compare results for different thresholds in the following.

\subsection{Red Sea surface temperature data}\label{subsec:redsea}
The surface temperature dataset comprises daily observations for the years 1985 to 2015 for 16703  locations across the Red Sea, corresponding to $0.05\degree\times0.05\degree$ grid cells. We focus only on the months of July to September to approximately eliminate the effects of seasonality. More information on the data, which were obtained from a combination of satellite and in situ observations, can be found in \cite{Donlon.etal.2012}. Extreme events in this dataset are of interest, since particularly high water temperatures can be detrimental to marine life, e.g., causing coral bleaching, and in some cases coral mortality.

\begin{figure}
    \centering
    \includegraphics[width=0.8\textwidth]{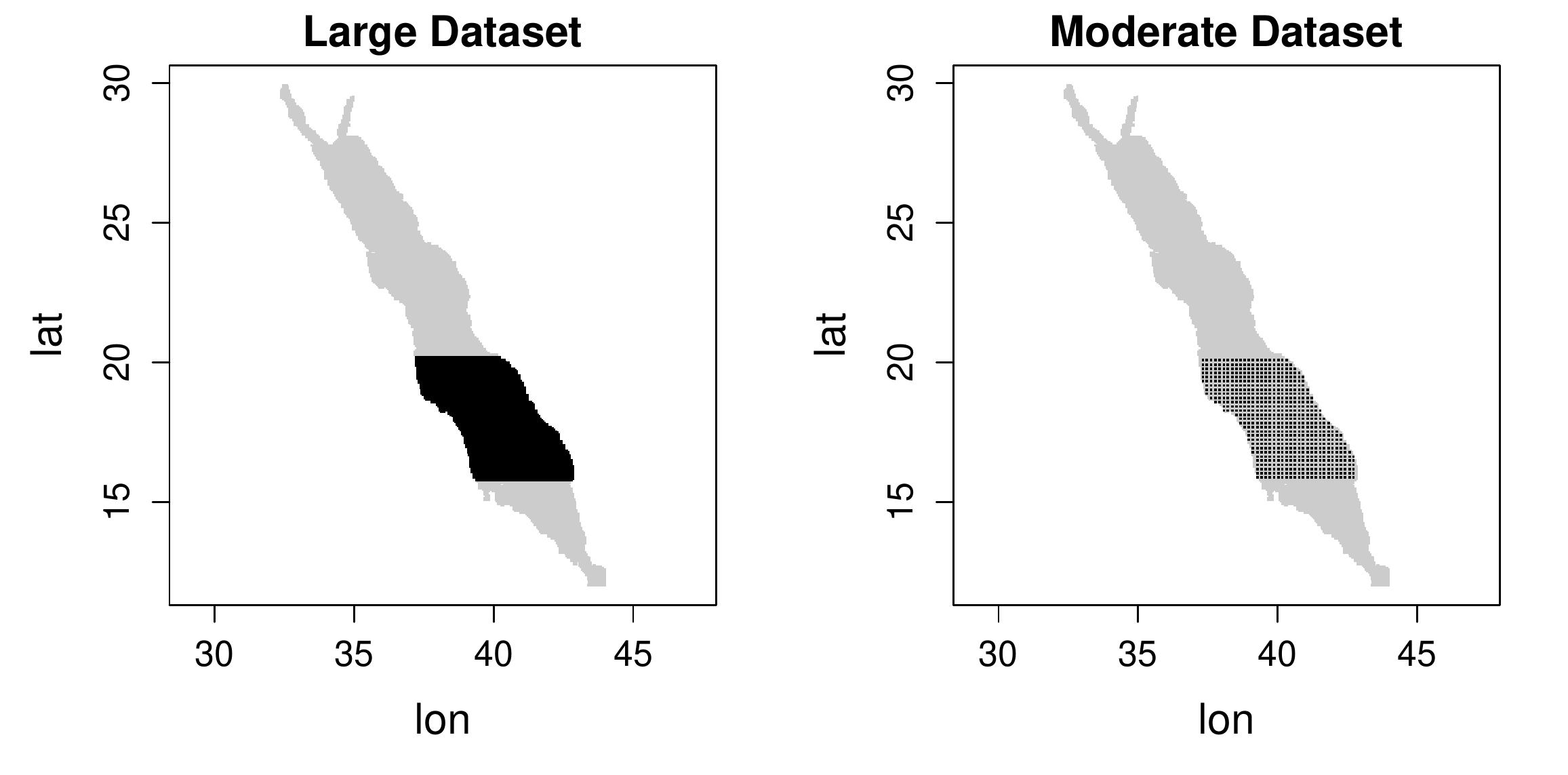}
    \caption{Location of the Red Sea (grey), and the subsets of grid cells in the two datasets we consider (black).}
    \label{fig:spatialLocations}
\end{figure}

\cite{Simpson.Wadsworth.2020} apply their conditional spatio-temporal extremes model to a subset of 54 grid cells located across the north of the Red Sea. In this paper, we instead focus on a southern portion of the Red Sea, where coral bleaching is currently more of a concern \citep{Fine.etal.2019}. We demonstrate our approach using datasets of two different spatial dimensions; the first dataset contains 6239 grid cells, corresponding to all available locations in our region of interest, while the second dataset is obtained by taking locations at every third longitude and latitude value in this region, leaving 678 grid cells to consider. These two sets of spatial locations are shown in Figure~\ref{fig:spatialLocations}. \cite{Simpson.Wadsworth.2020} consider their 54 spatial locations at five time-points, resulting in a lower dimensional problem than both the datasets we consider here. On the other hand, \cite{Hazra.Huser.2020} model the full set of 16703 grid cells, but they ensure computational feasibility by implementing so-called `low-rank' modeling techniques using spatial basis functions given by the dominant  empirical orthogonal functions, obtained from preliminary empirical estimation of the covariance matrix of the data.

\edit{There are two transformations that we apply to these data as a preliminary step.} First, as our study region lies away from the equator, one degree in latitude and one degree in longitude correspond to different distances. To correct for this we apply a transformation, multiplying the longitude and latitude values by 1.04 and 1.11, respectively, such that spatial coordinates are expressed in units of approximately $100$ km. \edit{Our resulting spatial domain measures approximately $400$ km between the east and west coasts and $500$ km from north to south.} We use the Euclidean distance on these transformed coordinates to measure spatial distances in the remainder of the paper. \edit{We also transform the margins to Laplace scale using the approach outlined in Section~\ref{subsec:marginalTransform}. We take $v$ in~\eqref{eqn:marginalDist} to be the empirical 0.95 quantile of $Y$, here representing the observed sea surface temperature at an individual location, so that $\lambda_v=0.05$.} \edit{Any temporal trend in the marginal distributions could also be accounted for at this stage, e.g., using the approach of \cite{Eastoe.Tawn.2009}, but we found no clear evidence that this was necessary in our case.}

\edit{In the remainder of this section, we apply a variety of conditional spatial extremes models to the Red Sea data with the large spatial dataset in Figure~\ref{fig:spatialLocations}, and apply several model selection and diagnostic techniques. For the conditioning site, we select a location lying towards the centre of the spatial domain of interest.} We discuss this choice further in Section~\ref{sec:sensitivity}, where we present additional results based on the moderate dataset. In Appendix~\ref{app:chi}, we provide an initial assessment of the spatial extremal dependence properties of the sea surface temperature data, based on the tail correlation function defined in~\eqref{eq:chi}. These results demonstrate that there is weakening dependence in the data at increasingly extreme levels, \edit{which provides an initial indication} that models exhibiting asymptotic independence should be more appropriate here.

\subsection{\edit{The mesh and prior distributions for the Red Sea data application}}\label{subsec:RedSeaMeshPriors}
\begin{figure}
    \centering
    \includegraphics[width=0.5\textwidth]{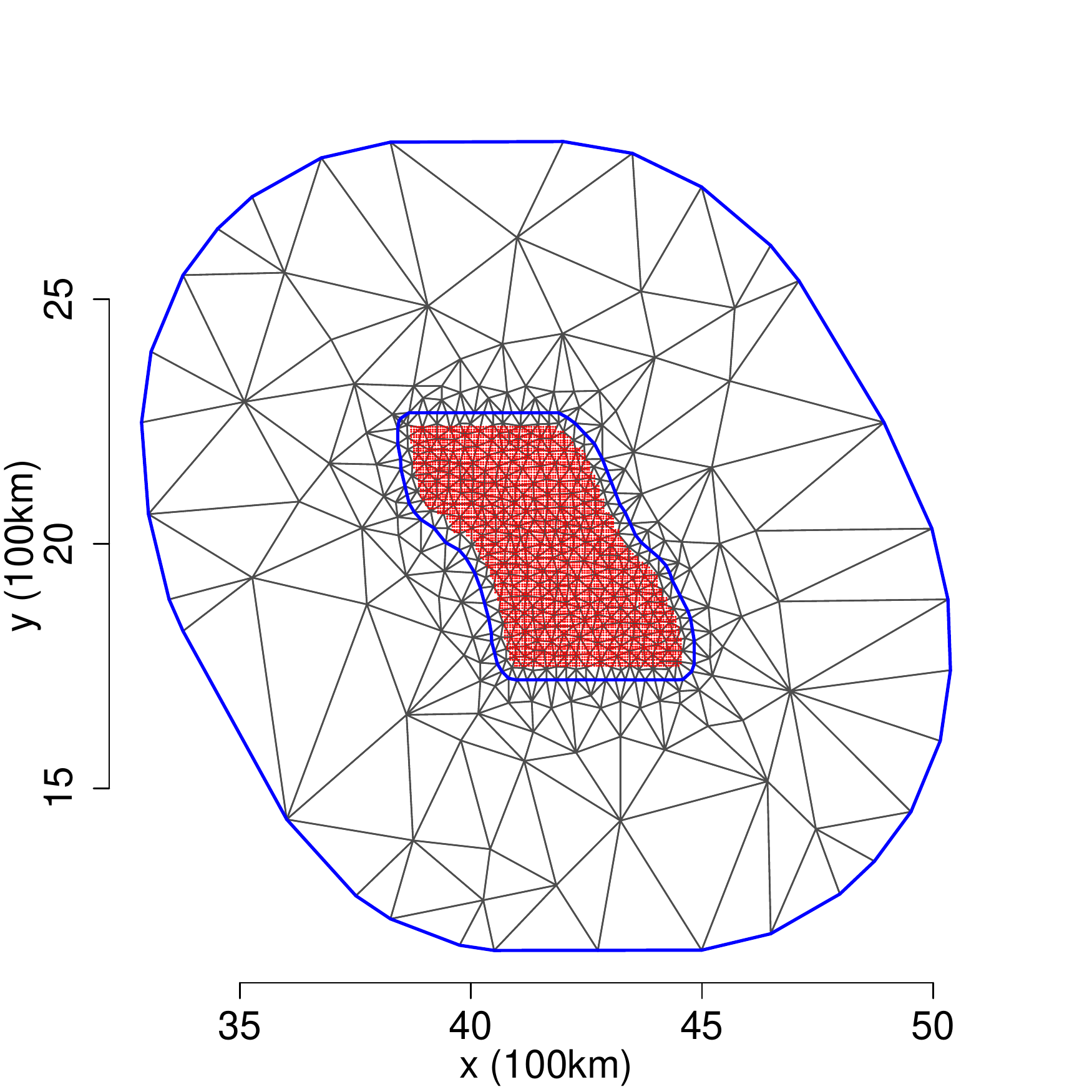}
    \caption{The large set of spatial locations (red dots) and the corresponding triangulation mesh used for the SPDE model with an inner and outer boundary (blue lines). The inner boundary delimits a high-resolution zone covering the study area, while the outer boundary delimits an extension zone with lower resolution to prevent effects of SPDE boundary conditions in the study area.}
    \label{fig:largeSpatialMesh}
\end{figure}

\edit{As discussed in Section~\ref{sec:hyperprior}, in order to carry out our latent variable approach to model inference, we require a triangulation of the area of interest. The mesh we use for the spatial domain in the southern Red Sea is shown in Figure~\ref{fig:largeSpatialMesh}. This was generated using \texttt{R-INLA}, with the most dense region corresponding to the area where we have observations. The spatial triangulation mesh has $541$ nodes, i.e., the dimension of the latent process is approximately 8.7\% of the size of the large dataset, and similar in size to the moderate dataset.} {In this case, the extension of the mesh beyond the study region is reasonably big, in order to avoid boundary effects of the SPDE for the sea surface temperatures, whose spatial dependence range is known to be relatively large; see \citet{Simpson.Wadsworth.2020}.} We use a coarser resolution in the extension region to keep the number of latent Gaussian variables as small as reasonably possible.

Due to the availability of many observations in the Red Sea dataset, we found the hyperparameter priors to only have a small impact on posterior inference in our application, and that the credible intervals of the hyperparameters are very narrow. We have chosen moderately informative PC priors through the following specification:
$$
\mathrm{Pr}(\sigma > 0.1)=0.5, \quad \mathrm{Pr}(\sigma_{Z} > 1) = 0.5, \quad  \mathrm{Pr}(\rho_{Z} > 100\ \mathrm{km})=0.5,
$$
where $\sigma_{Z}$ and $\rho_{Z}$ are the standard deviation and the empirical range, respectively, of the unconstrained spatial Mat\'ern fields $\{Z(s)\}$. \edit{Where the $\beta$-parameter is to be estimated as part of a specified model, we choose a log-normal prior where the normal has mean $-\log(2)$ and variance $1$. This does not guarantee estimates of $\beta<1$, but such a constraint could be included within the \texttt{generic}-model framework if required.} \edit{The Mat\'ern covariance function is also specified by a shape parameter $\nu$. We fix $\nu=0.5$, corresponding to an exponential correlation function. Sensitivity to $\nu$} can be checked by comparing fitted models across different values, as demonstrated in Appendix~\ref{app:zetaSensitivity}. We find this to have little effect on the results for our data. 

\edit{As discussed in Section~\ref{sec:hyperprior}, for each spline function, we place one knot at the boundary where $s=s_0$ and use a further 14 equidistant interior knots.} This quantity provides a reasonable balance between the reduced flexibility that occurs when using too few knots, and the computational cost and numerical instability (owing to near singular precision matrices) that may arise with using too many. \edit{For these spline components, we have fixed the range to $100$~km and the standard deviation to $0.5$. If we wished to obtain very smooth posterior estimates of the spline function, we could choose parameters that lead to stronger constraints on the (prior) variability of the spline curve.} We will demonstrate the estimated spline functions for some of our models in Section~\ref{subsec:Model3}.

\subsection{Variants of the spatial model and model comparison}\label{subsec:spatialModels}
\edit{In Section~\ref{sec:proposed.a}, we discussed choices of the normalizing functions $a_{s-s_0}(x)$ and $b_{s-s_0}(x)$ that are possible under the INLA framework. In Table~\ref{tab:spatialModels}, we summarize the models we consider based on the structures outlined in equation~\eqref{eq:modelgeneral}. We fit models with these different forms and subsequently select the best model for our data. Model~0 has $\alpha(s-s_0)=1$, $\beta=0$, resulting in a very simple asymptotically dependent model.} Model~2 is also asymptotically dependent, but allows for weaker dependence than in Model 0 due to the drift  that is captured by the $\gamma(s-s_0)$ term, supposed to be negative in practice. In Model~6, the residual process has been removed, so that all variability is forced to be captured by the term $\epsilon_i$ in \eqref{eq:modelgeneral}. Models~0~and~6 are meant to act as simple baselines to which we can compare the other models, but we would not expect them to perform sufficiently well in practice. \edit{As mentioned in Section~\ref{subsec:CSEintro}, \cite{Wadsworth.Tawn.2019} propose two options for constructing the residual process $\{Z^0(s)\}$, each based on manipulation of a stationary Gaussian process $\{Z(s)\}$. For all models in Table~\ref{tab:spatialModels}, we focus on a residual process of the form $\{Z^0(s)\}=\{Z(s)\}-Z(s_0)$, further detail on which is given in Section~\ref{subsec:residualConstraint}.} 

\begin{table}[ht]
\centering
 \begin{tabular}{|c|c|c c c|c|} 
 \hline
 Model number & Model form & WAIC & CPO & RMSE & Run-time\\ 
 \hline
 0 
 & $x + \{Z^0(s)\}$ & 2438 & -0.0061 & 0.019 & 20\\
 1 
 & $x\cdot\alpha(s-s_0) + \{Z^0(s)\}$ & 614 &  -0.0028 & 0.001 & 22\\  
 2 
 & $x + \gamma(s-s_0) + \{Z^0(s)\}$ & 743 & -0.0035 & 0.005 & 35\\  
 3 
 & $x\cdot\alpha(s-s_0) + \gamma(s-s_0) + \{Z^0(s)\}$ & 4 & -0.0018 & 0 & 32\\  
 4 
 & $x\cdot\alpha(s-s_0) + \gamma(s-s_0) + x^\beta \cdot \{Z^0(s)\}$ &  0  & 0 & 0.003 & 107\\ 
 5 
 & $x\cdot\alpha(s-s_0) + x^\beta \cdot \{Z^0(s)\}$ &  611 & -0.0004 & 0.010 & 86\\ 
 6 
 & $x\cdot\alpha(s-s_0) + \gamma(s-s_0)$ & 4394961 & -2.8042 & 0.514 & 43\\ 
 \hline
\end{tabular}
 \caption{Summary of conditional spatial models, model selection criteria, and total run-times (minutes). The minimum WAIC value (-1460982 for Model~4); maximum CPO value (3.0305 for Model~4); and minimum RMSE value (0.862 for Model~3) have been subtracted from their respective columns. We estimate $\beta$ as 0.29 with a 95\% credible interval of $(0.27,0.31)$ (Model~4) and 0.33 $(0.31,0.34)$ (Model~5).}
 \label{tab:spatialModels}  
\end{table}

\edit{Alongside the models in Table~\ref{tab:spatialModels}, we provide the corresponding WAIC and CPO values, as discussed in Sections~\ref{subsec:spatialModelSelection}~and~\ref{subsec:loocv}, respectively. The computation times for each model are also included, as this information may also aid model selection where there is similar performance under the other criteria.} 

\edit{Beginning with the WAIC, we first recall that smaller values of this criterion are preferred. Models~1~and~3 are simplified versions of Models~5~and~4, respectively, in that the value of $\beta$ is fixed to 0 rather than estimated directly in \texttt{R-INLA}. In both cases, the results are very similar whether $\beta$ is estimated or fixed, suggesting the simpler models with $\beta=0$ are still effective. The estimated WAIC values suggest Models~3~and~4 provide the best fit for our data. The common structure in these models are the terms $\alpha(s-s_0)x$ and $\gamma(s-s_0)$, indicating that their inclusion is important. In Model~4, the posterior mean estimate of $\beta$ is 0.29, but despite simplifying the model, setting $\beta=0$ as in Model~3 provides competitive results.}

\edit{On the other hand, the CPO results are relatively similar across Models 0~to~5, but clearly substantially better than Model~6, which we include purely for comparison. Model~6 performs poorly here since all spatially correlated residual variation has been removed. We provide a histogram of PIT values for Model~3 in Appendix~\ref{app:pit}, with equivalent plots for Models~0~to~5 being very similar. The histogram has a peak in the middle, suggesting that the posterior predictive densities for single observations are generally slightly too ``flat''; however, here the variability in the posterior predictive distributions is very small throughout. Therefore, the fact of slightly overestimating the true variability, which is very small, does not cause too much concern about the model fit.} If the PIT values concentrated strongly at $0$ and $1$, this would indicate that posterior predictive distributions would not allow for enough uncertainty, i.e., the model would be overconfident with its predictions; however, this is not the case here.  Due to the smoothness of our data we essentially have perfect predictions using Models~0~to~5, and these plots are not particularly informative, but again may be useful in settings where spatio-temporal interpolation is a modeling goal.

\edit{Finally, we consider using a further cross validation procedure to compare the different models.} This involves removing all data for locations lying in a quadrant to the south-east of the conditioning site, and using our methods to estimate these values. The difference between the estimates and original data, on the Laplace scale, can be summarized using the root mean square error (RMSE). These results are also provided in Table~\ref{tab:spatialModels}, where Model~3 gives the best results, although is only slightly favoured over the other Models~0~to~5.

\subsection{Results for Model~3}\label{subsec:Model3}
For our application, it is difficult to distinguish between the performance of Models 0~to~5 using the cross validation approaches, but Models~3~and~4 both perform well in terms of the WAIC. We note that the run-time for Model~3, provided in Table~\ref{tab:spatialModels}, is less than one third of the run-time of Model~4 for this data, so we choose to focus on Model~3 here due to its simplicity. In general, simpler models have quicker computation times, but this is not necessarily always the case; we comment further on this in Section~\ref{subsec:space-time-diagnostics}. We provide a summary of the fitted model parameters for Model~3, excluding the spline functions, in Table~\ref{tab:parameterEstimates}. The estimated value of $\sigma^2$ is very small, as expected. The Mat\'{e}rn covariance of the process $\{Z(s)\}$ has a reasonably large dependence range, estimated to be 428.2~km. 

\begin{table}[ht]
\centering
 \begin{tabular}{|c|c c|} 
 \hline
 Parameter & Posterior mean & 95\% credible interval \\ 
 \hline
  $\sigma^2$ & 0.0107 & (0.0106, 0.0107) \\
  $\sigma_Z$ & 1.557 & (1.496, 1.618) \\
  $\rho_Z$ & 428.2 & (409.5, 446.8) \\
 \hline
 \end{tabular}
 \caption{Estimated parameters for Model 3.}
 \label{tab:parameterEstimates}
\end{table}

\begin{figure}[!htbp]
    \centering
    \includegraphics[width=0.9\textwidth]{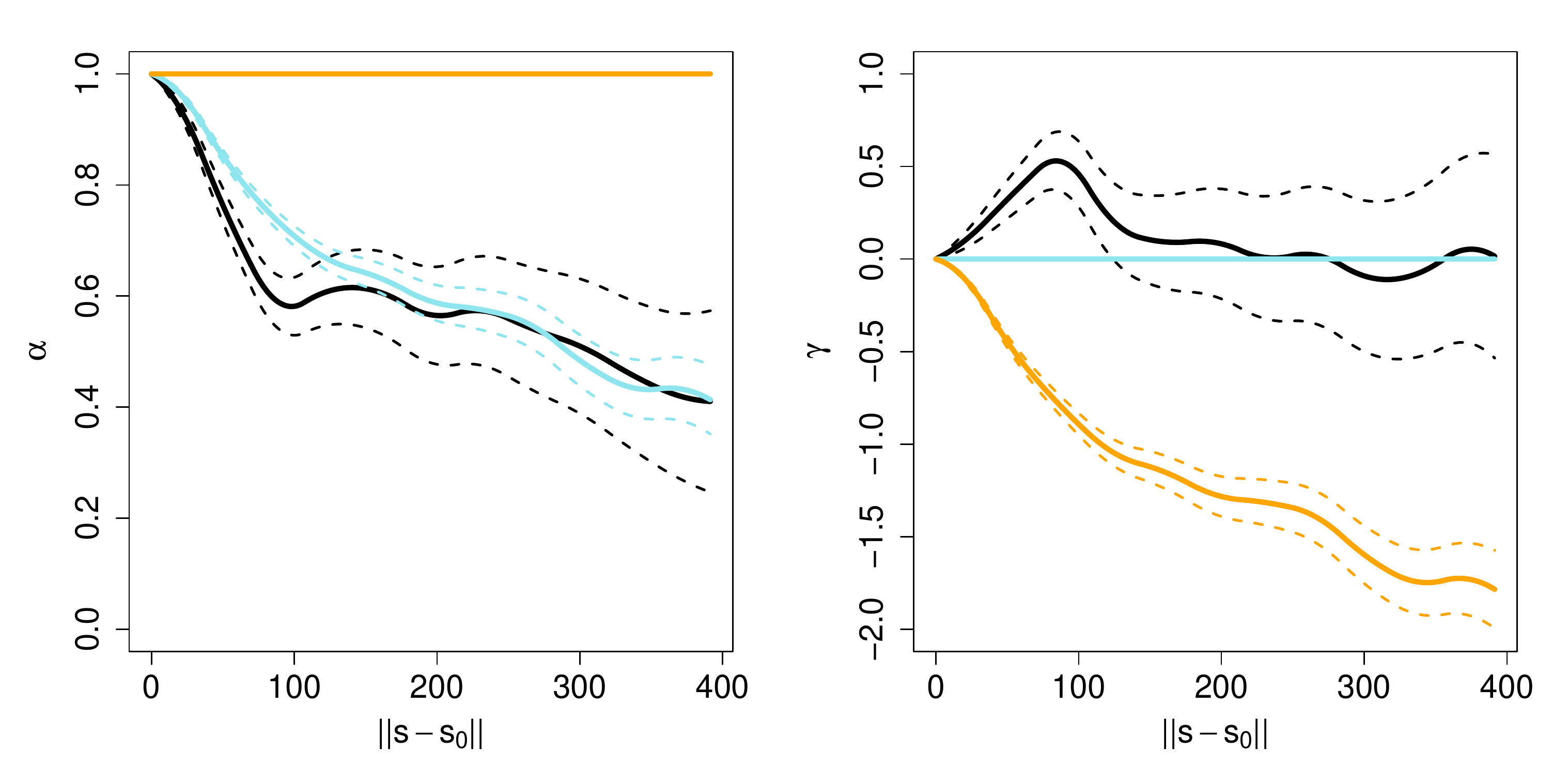}
    \caption{Posterior mean estimates of the spline functions $\alpha(s-s_0)$ (left) and $\gamma(s-s_0)$ (right) for Model~1 (blue), Model 2 (orange) and Model~3 (black). \edit{The dashed lines show approximate 95\% pointwise credible intervals in each case.}}
    \label{fig:model4splines}
    \vspace{1cm}\centering
    \includegraphics[width=0.9\textwidth]{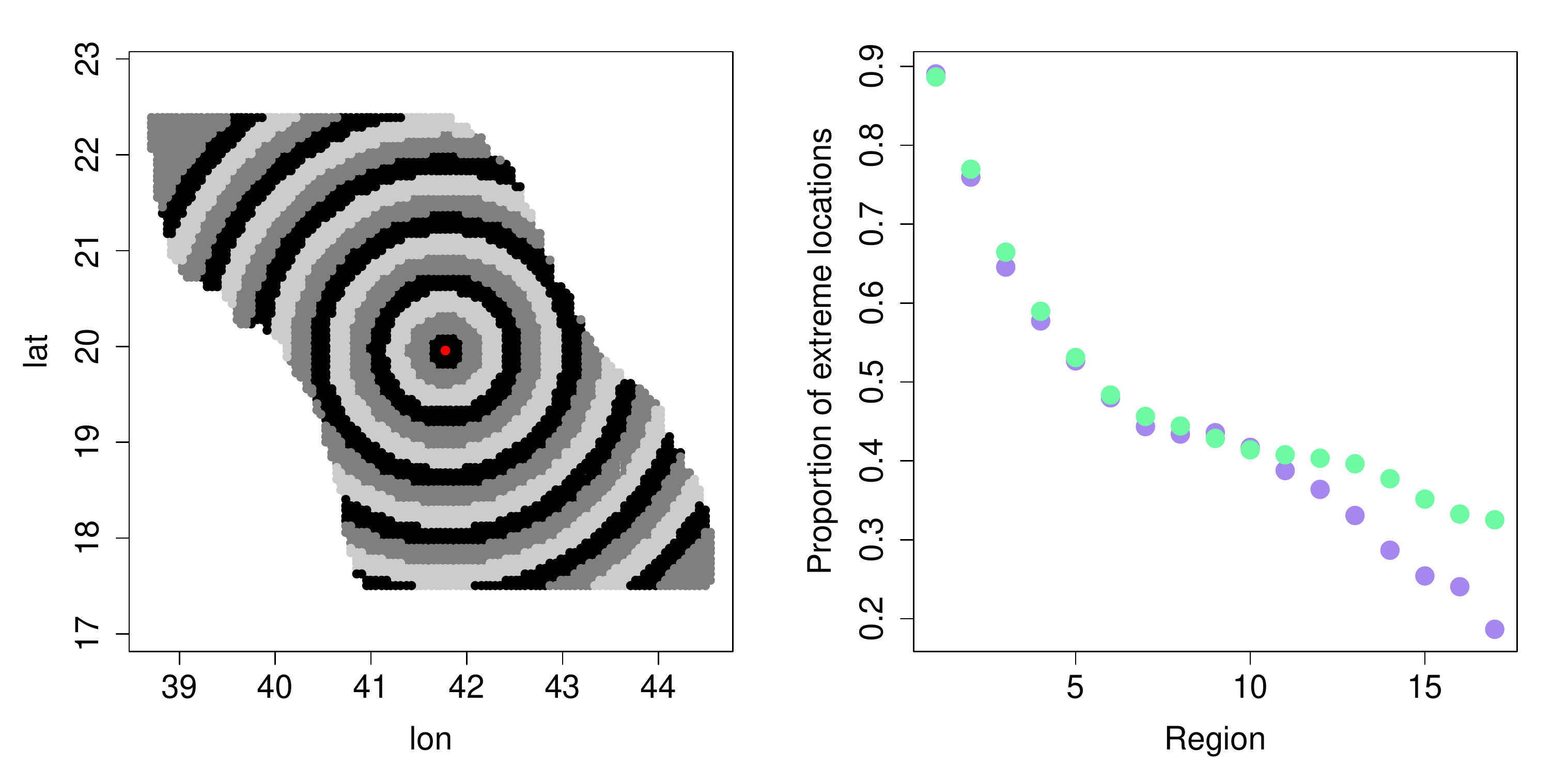}
    \caption{Left: the spatial domain separated into 17 regions; the region labels begin at 1 in the centre of the domain, and increase with distance from the centre. The conditioning site $s_0$ is shown in red. Right: the estimated proportion of locations that exceed the 0.95 quantile, given it is exceeded at $s_0$ using Model~3 (green) and equivalent empirical results (purple).}
    \label{fig:extremeProb1}
\end{figure}

We now consider the estimated spline functions $\alpha(s-s_0)$ and $\gamma(s-s_0)$ for Model~3; these are shown in Figure~\ref{fig:model4splines}. For comparison, we also show the estimate of $\alpha(s-s_0)$ for Model~1 and $\gamma(s-s_0)$ for Model~2. These two models are similar to Model~3, in that $\beta=0$, but $\gamma(s-s_0)=0$ for Model~1 and $\alpha(s-s_0)=1$ for Model~2, for all $s\in\mathcal{S}$. For Model~1, the  $\alpha(s-s_0)$ spline function generally decreases monotonically with distance, as would be expected in spatial modeling. For Model~3, the interaction between the two spline functions makes this feature harder to assess, but further investigations have shown that although $\alpha(s-s_0)$ and $\gamma(s-s_0)$ are not monotonic in form, the combination $\alpha(s-s_0) x+\gamma(s-s_0)$ is usually decreasing for $x\geq u$; i.e., there is posterior negative correlation, and transfer of information between the two spline functions. Some examples of this are given in Figure~\ref{fig:diffs0splines} and will be discussed in Section~\ref{sec:sensitivity}. The behaviour of the $\gamma(s-s_0)$ spline function for Model~2 is similar to that of the $\alpha(s-s_0)$ functions for Models~1~and~3, highlighting that all three models are able to capture similar features of the data despite their different forms. The success of Model~3 over Models~1~and~2 can be attributed to the additional flexibility obtained via the inclusion of both spline functions. We note that in terms of representing the data, there may be little difference between suitable models, as we see here. However, we should also consider the diagnostics relating to our specific modeling purpose, which in our case is extrapolation. The results in Appendix~\ref{app:chi} demonstrate that there is weakening dependence in the Red Sea data. The asymptotic dependence of Model~2 means that it cannot capture this feature, and is therefore unsuitable here. \edit{In Appendix~\ref{app:chi}, we compare the empirical tail correlation estimates to ones obtained using simulations from our fitted Model~3. This model goes some way to capturing the observed weakening dependence, although the estimates do not decrease as quickly as for the empirical results. We comment further on this issue below.}

Our fitted models can be used to obtain estimates of quantities relevant to the data application. For sea surface temperatures, we may be interested in the spatial extent of extreme events. High surface water temperatures can be an indicator of potentially damaging conditions for coral reefs, so it may be useful to determine how far-reaching such events could be. To consider such results, we fix the model hyperparameters and spline functions to their posterior means, and simulate directly  from the spatial residual process of Model~3. If a thorough assessment of the uncertainty in these estimates was required, we could take repeated samples from the posterior distributions of the model parameters fixed to their posterior mean, and use each of these to simulate from the model. However, assessing the predictive distribution in this way is computationally more expensive, so we proceed without this step.

We separate the spatial domain into the 17 regions demonstrated in the left panel of Figure~\ref{fig:extremeProb1}. Given that the value at the conditioning site exceeds the 0.95 quantile, we estimate the proportion of locations in each region that also exceed this quantile. Results obtained via 10,000 simulations from Model~3 are shown in the right panel of Figure~\ref{fig:extremeProb1}, alongside empirical estimates from the data. These results suggest that Model~3 provides a successful fit of the extreme events, particularly within the first ten regions, which correspond to distances up to approximately 200~km from the conditioning site. At longer distances, the results \edit{do differ, which may be due to the comparatively small number of locations that contribute to the model fit in these regions and to some mild non-stationarities arising close to the coastline}. \edit{In Appendix~\ref{app:extrapolationDiagnostic}, we present a similar diagnostic where we instead extrapolate to the 0.99 quantile. Comparing the empirical results to those in Figure~\ref{fig:extremeProb1}, we again see that the data exhibits weakening dependence as we increase the threshold level. This suggests that an asymptotically independent model, as we have with Model~3, is appropriate; an asymptotically dependent model would not have captured this feature. However, these diagnostics do suggest that the dependence does not weaken quickly enough in our fitted model. It is possible that this could have been improved by a different threshold choice, but investigating this is beyond the scope of the paper.}

\subsection{Sensitivity to the conditioning site}
\label{sec:sensitivity}

A natural question when applying the conditional approach to spatial extreme value modeling, is how to select the conditioning location. Under an assumption of spatial stationarity in the dependence structure, the parameters of the conditional model defined in~\eqref{eqn:modelingAssumption} should be the same regardless of the location $s_0$. However, since the data are used in slightly different ways for each conditioning site, and because the stationarity assumption is rarely perfect, we can expect some variation in parameter estimates for different choices of $s_0$.

In \citet{Wadsworth.Tawn.2019} and \citet{Simpson.Wadsworth.2020}, this issue was circumvented by using a composite likelihood that combined all possible individual likelihoods for each conditioning site, leading to estimation of a single set of parameters that reduced sampling variability and represented the data well on average. However, bootstrap methods are needed to assess parameter uncertainty, and as the composite likelihood is formed from the product of $d$ full likelihoods, the approach scales poorly with the number of conditioning sites. Composite likelihoods do not tie in naturally to Bayesian inference as facilitated by the INLA framework, and so to keep the scalability, and straightforward interpretation of parameter uncertainty, we focus on implementations with a single conditioning location. Sensitivity to the particular location can be assessed similarly to other modeling choices, such as the threshold above which the model is fitted.

\begin{figure}
    \centering
    \includegraphics[width=\textwidth]{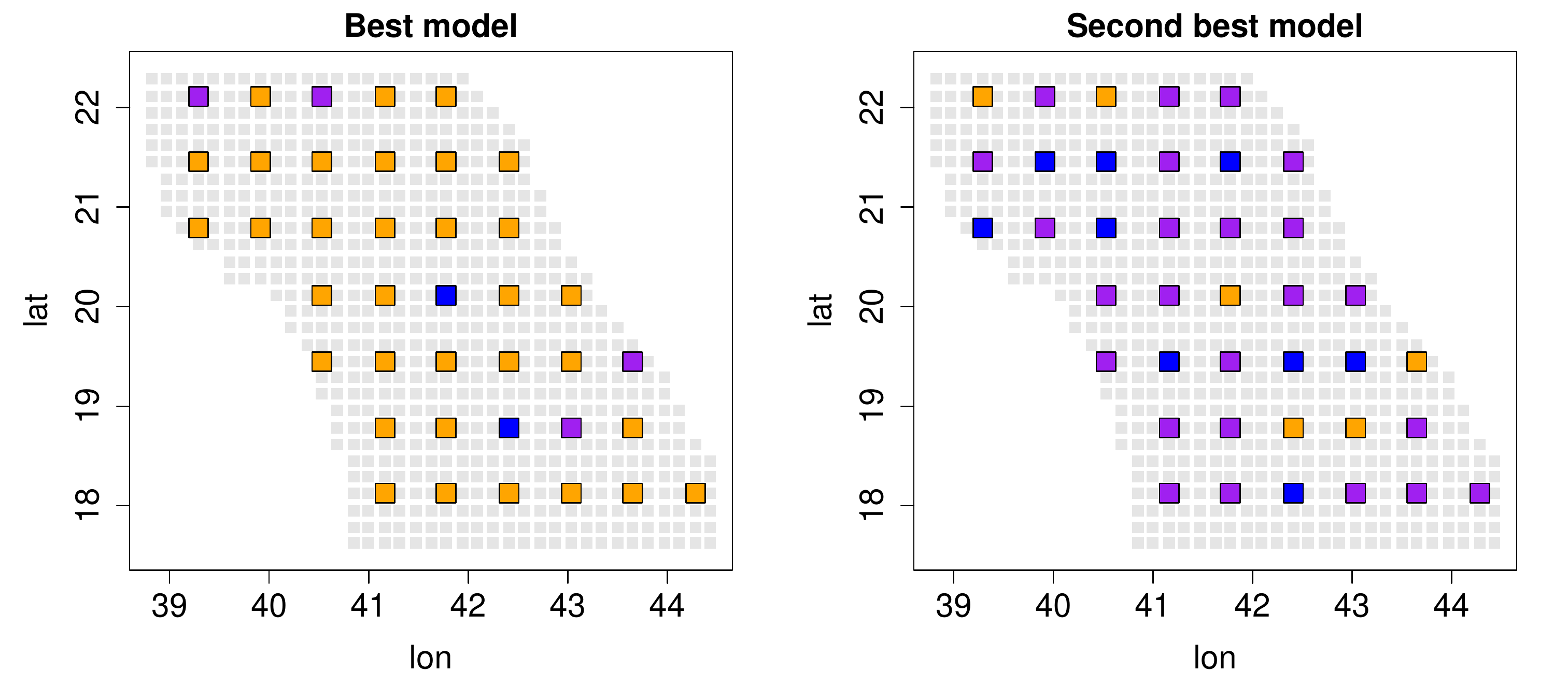}
    \caption{Maps showing the `best' and `second best' models using different conditioning sites, based on minimizing the WAIC: Model~3, blue; Model~4, orange; Model~5, purple.}
    \label{fig:diffs0results}
    \vspace{1cm}
    \includegraphics[width=\textwidth]{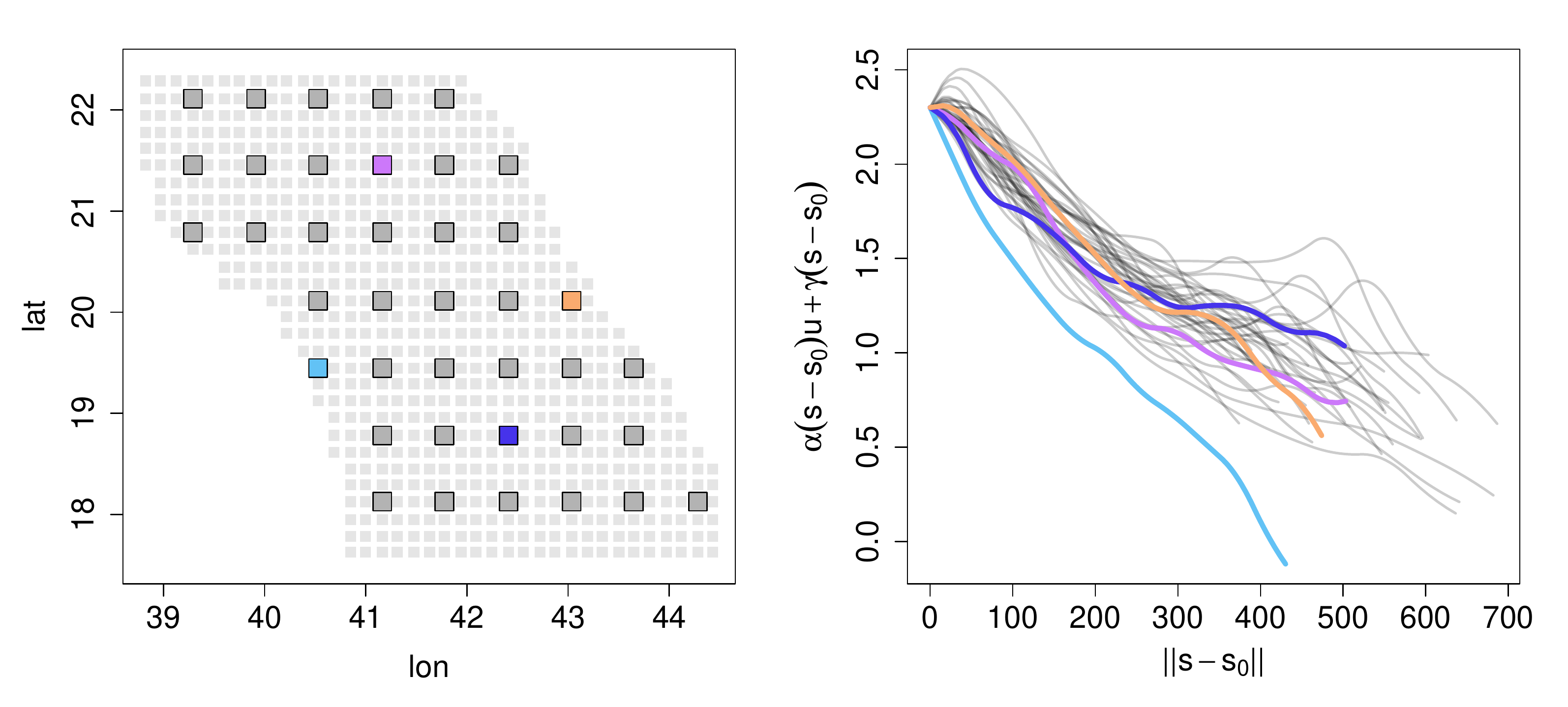}
    \caption{\edit{Posterior mean estimates of $\alpha(s-s_0)u + \gamma(s-s_0)$} for Model~3 (right), with $u$ representing the threshold used in the model fits. The colours of the lines correspond to the conditioning sites used, as shown in the left panel.}
    \label{fig:diffs0splines}
\end{figure}

In particular, different conditioning sites may lead us to select different forms of the models described in Table~\ref{tab:spatialModels}, as well as the resulting parameter estimates. To assess this, we fit all seven models to the moderate dataset, using 39 different conditioning sites on a grid across the spatial domain, with the mesh and prior distributions selected as previously. We compare the models using the WAIC, as described in Section~\ref{subsec:spatialModelSelection}.
The results are shown in Figure~\ref{fig:diffs0results}, where we demonstrate the best two models for each conditioning site. For the majority of cases, Model~4 performs the best in terms of the WAIC, and in fact, it is in the top two best-performing models for all conditioning sites. The best two performing models are either Models~3~and~4 or Models~4~and~5 for all conditioning locations. This demonstrates that there is reasonable agreement across the spatial domain, and suggests that using just one conditioning location should not cause an issue in terms of model selection.

To further consider how restrictive it is to only fit models at one conditioning site, we can compare the spline functions estimated using different locations for $s_0$. We again focus on results for Model~3, as in Section~\ref{subsec:Model3}, and consider estimates of $\alpha(s-s_0)u + \gamma(s-s_0)$, with $u$ representing the threshold used for fitting. We demonstrate the estimates of this function in Figure~\ref{fig:diffs0splines}, for the same 39 conditioning sites used in Figure~\ref{fig:diffs0results}, highlighting results for four of these sites situated across the spatial domain. Overall, the estimated functions are reasonably similar, particularly for shorter distances. There is one function that appears to be an outlier, corresponding to a conditioning site located on the coast. Although the other coastal conditioning sites we consider do not have this issue, it does suggest that some care should be taken here.

\begin{figure}[!htbp]
    \centering
    \includegraphics[width=0.9\textwidth]{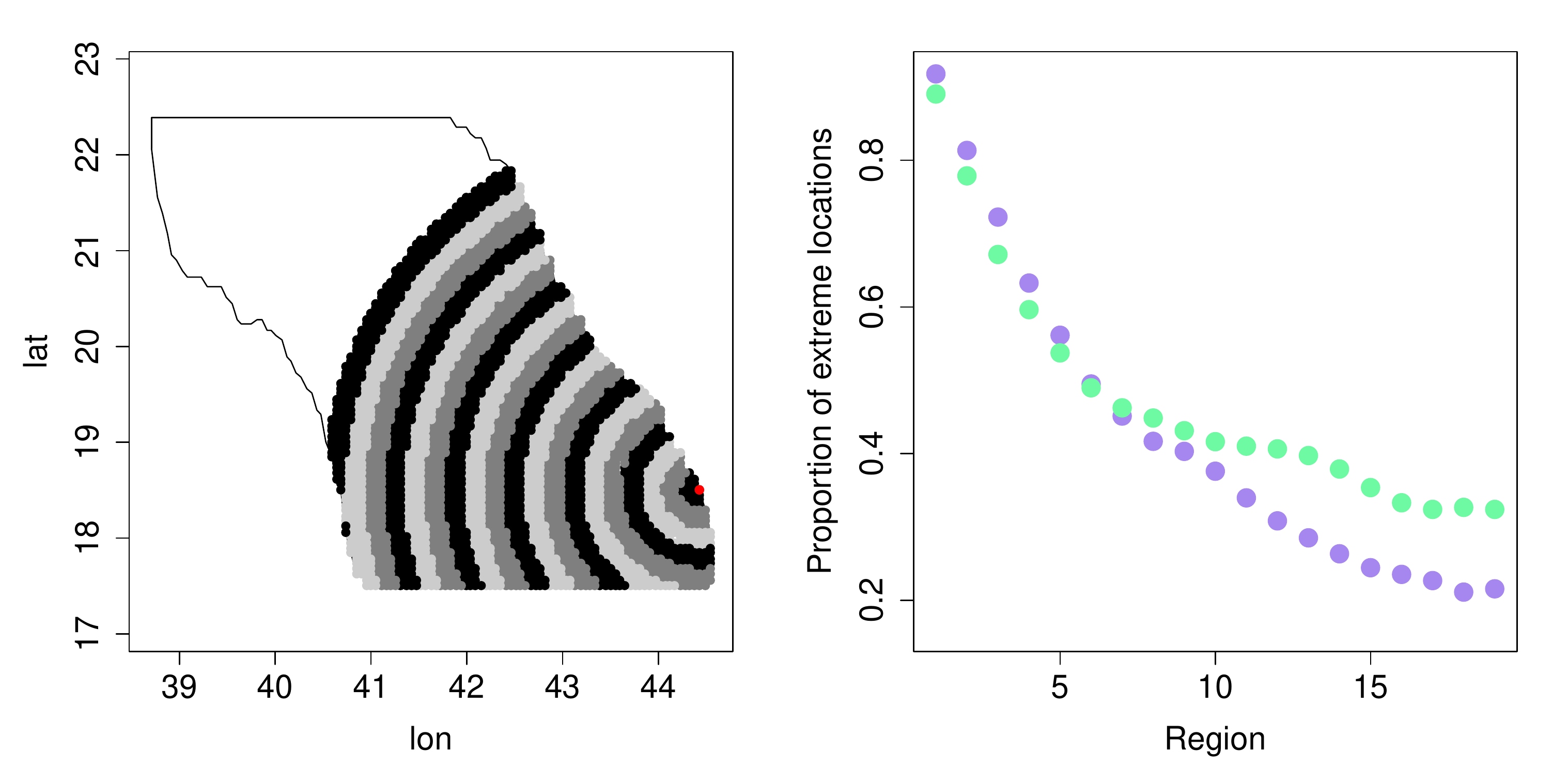}
    \caption{Left: the spatial domain separated into regions; the region labels begin at 1 at $s_0$ (red), and increase with distance from this location. Right: the estimated proportion of locations that exceed the 0.95 quantile, given it is exceeded at $s_0$ using Model~3 (green) and equivalent empirical results (purple).}
    \label{fig:extremeProb_diffs0}
\end{figure}

As a final test on the sensitivity to the conditioning site, we consider the implications if we fit Model~3 at one conditioning site, and use this for inference at another location. In particular, we take the results from Section~\ref{subsec:Model3}, using a conditioning site near the centre of the spatial domain, and use these to make inference at a conditioning site located on the coast. We use a method analogous to the one used to create Figure~\ref{fig:extremeProb1}. That is, we separate the spatial domain into regions, and for each one, we estimate the proportion of locations that take values above their 0.95 quantile, given that this quantile is exceeded at the conditioning location. In Figure~\ref{fig:extremeProb_diffs0}, we compare results based on simulations from the fitted model to empirical estimates. Although a different conditioning location was used to obtain the model fit, the results are still good, particularly up to moderate distances, supporting our use of a single conditioning site for inference. One issue that is highlighted here is that by fitting the model at a central conditioning site, the maximum distance to $s_0$ is around 391~km, so we are not able to make inference about the full domain for a conditioning site near the boundary, where the maximum distance to other locations is much larger. This aspect should be taken into account when choosing a conditioning site for inference. This issue is specific to the use of spline functions for $\alpha(s-s_0)$ and $\gamma(s-s_0)$, and there is no such problem for parametric functions such as the one proposed by \cite{Wadsworth.Tawn.2019} for $\alpha(s-s_0)$. There is therefore a trade-off here between the flexbility of the splines and the spatial extrapolation possible using parametric functions.

\section{Inference for conditional space-time extremes}\label{sec:spacetimeInference}

\subsection{Conditional spatio-temporal extremes models}

\cite{Simpson.Wadsworth.2020} extend assumption~\eqref{eqn:CSEassumption} to a spatio-temporal setting. The aim is to model the stationary process $\{X(s,t):(s,t)\in\mathcal{S}\times\mathcal{T}\}$ which also has marginal distributions with exponential upper tails. The conditioning site is now taken to be a single observed space-time location $(s_0,t_0)$, and the model is constructed for a finite number of points $(s_1,t_1),(s_1,t_2),\dots,(s_d,t_\ell)$ pertaining to the process at $d$ spatial locations and $\ell$ points in time, where data may be missing for some of the space-time points. The structure of the conditional extremes assumption is very similar to the spatial case, in particular, it is assumed that there exist functions $a_{(s,t)-(s_0,t_0)}(\cdot)$ and $b_{(s,t)-(s_0,t_0)}(\cdot)$ such that as $u\rightarrow\infty$,
$$
\Pr\left(\left[\frac{X(s_i,t_j)-a_{(s_i,t_j)-(s_0,t_0)}\left\{X(s_0,t_0)\right\}}{b_{(s_i,t_j)-(s_0,t_0)}\left\{X(s_0,t_0)\right\}}\right]_{\substack{i=1,\dots,d,\\j=1,\dots,\ell}} \leq \bm{z}~\bigg{\vert}~X(s_0,t_0)=u\right) ~\to \Pr\left[\{Z^0(s_i,t_j)\}_{\substack{i=1,\dots,d,\\j=1,\dots,\ell}} \leq \bm{z}\right],
$$
for \edit{$\{Z^0(s_i,t_j)\}_{\substack{i=1,\dots,d,\\j=1,\dots,\ell}}$ representing finite-dimensional realizations of} a spatio-temporal residual process $\{Z^0(s,t)\}$. Once more the excesses $X(s_0,t_0)-u|X(s_0,t_0)>u$ are independent of the residual process as $u \to \infty$, and the constraints on the residual process $\{Z^0(s,t)\}$ and normalizing function $a_{(s,t)-(s_0,t_0)}(\cdot)$ are analogous to the spatial case. We consider spatio-temporal variants of spatial models 1, 3, 4 and 5, which provided the best WAIC values, in Section~\ref{subsec:spatialModels}; see the model summary in Table~\ref{tab:spacetimeResults}. In order to preserve sparsity in the precision matrix \edit{of the relevant latent variables}, a simple autoregressive structure is employed for the temporal aspect of the residual process; further details are provided in Section~\ref{subsec:spacetime}. Specifically, we construct the process $\{Z^0(s,t)\}$ as $\{Z(s,t)\}-Z(s_0,t_0)$ using the first-order autoregressive structure in combination with the spatial SPDE model as described in equation~\eqref{eq:ar1}. \edit{For the temporal auto-correlation coefficient $\rho$ in~\eqref{eq:ar1}, we again opt for a PC prior. The baseline could be either $\rho=0$ (no dependence) or $\rho=1$ (full dependence); here, we choose $\rho=0$ and a moderately informative prior through the specification $\mathrm{Pr}(\rho > 0.5) = 0.5$.} The prior distributions for $\alpha(s-s_0,t-t_0)$ and $\gamma(s-s_0,t-t_0)$ are constructed according to \eqref{eq:ar1}, with a \edit{one-dimensional} SPDE model for a quadratic spline with $14$ interior knots deployed for spatial distance \edit{and replicated for each of the $\ell$ time points, with prior temporal dependence of spline coefficients for consecutive time lags controlled by a first-order autoregressive structure; the resulting Gaussian prior processes} are conditioned to have $\alpha(0,0)=1$ and $\gamma(0,0)=0$. \edit{Contrary to $Z^0$, the components $\alpha$ and $\gamma$ are deterministic in the conditional extremes framework, but through the semiparametric formulation we can handle them in the same way within the INLA framework. Using Gaussian process priors for spline coefficients allows for high modeling flexiblity through a relatively large number of basis functions, where hyperparameters ensure an appropriate smoothness of estimated functions.}   

\subsection{Spatio-temporal Red Sea surface temperature data}\label{subsec:RedSea_spacetime}
Since the spatio-temporal models are more computationally intensive than their spatial counterparts due to a larger number of hyperparameters and more complex precision matrices, we focus only on the moderate set of spatial locations demonstrated in Figure~\ref{fig:spatialLocations}, which contains 678 spatial locations; this will still result in a substantial number of dimensions when we also take the temporal aspect into account.

To carry out inference for the conditional spatio-temporal model, \edit{we must separate the data into temporal blocks of equal length, with the aim that each block corresponds to an independent observation from the process $\{X(s,t)\}$. We first apply a version of the runs method of \cite{Smith.Weissman.1994} to decluster the data. Each cluster corresponds to a series of observations starting and ending with an exceedance of the threshold $u$ at the conditioning site, with clusters separated by at least $r$ non-exceedances of $u$. Once these clusters are obtained, we take the first observation in each one as the start of an extreme episode, with the following six days making up the rest of the block. Declustering is applied only with respect to the spatial conditioning site $s_0$, but we still consider observations across all spatial locations at the corresponding time-points. We select the tuning parameter in the runs method to be $r=12$; this is chosen following the approach of \cite{Simpson.Wadsworth.2020}, who check for stability in the number of clusters obtained using different values of $r$,} and note that since we focus only on summer months, blocks should not be allowed to span multiple years. This declustering approach yields $28$ \edit{non-overlapping} blocks of seven days to which we can apply our four spatio-temporal models.

\subsection{Model selection, forecasting and cross validation}\label{subsec:space-time-diagnostics}
We compare the four models using similar criteria as in the spatial case. The WAIC and average CPO values are presented in Table~\ref{tab:spacetimeResults}, where the most complex Model~4 performs best in terms of the WAIC, while a slightly better CPO value arises for Model~3. We note that the model selected using the WAIC has the same form in both the spatial and spatio-temporal cases.

We also compare fitted and observed values using a variant of the root mean square error (RMSE). The results of within-sample RMSEs are almost identical for the four models, and therefore not included in Table~\ref{tab:spacetimeResults}, yielding a value of $0.077$. To assess predictive performance, it is more interesting to consider an additional variant of cross validation in the spatio-temporal case to test the forecasting ability of the models. We carry out seven-fold cross validation by randomly separating our 28 declustered blocks into groups of four, and for each of these groups we remove the observations at all locations for days two to seven. We then fit the model using the remaining data in each case, and obtain predictions  for the data that have been removed. This cross validation procedure is straightforward to implement, as in \texttt{R-INLA} it is possible to obtain predictions (e.g., posterior predictive means), including for time-points or spatial locations without observations. We compare the predicted values with the observations that were previously removed, presenting the cross validation root mean square error (RMSE$_{\text{CV}}$) in Table~\ref{tab:spacetimeResults}. Again, the results are quite similar, but Model~3 performs slightly better than the others. Finally, run-times are reported in the table and range between 1 hour and 4 hours, using 2 cores on machines with 32Gb of memory.
When comparing the spatial models with the corresponding space-time models having the same spatial component, the order of run-times  changes in our results. We emphasize that, on average, more complex latent models will require longer run-times with INLA if the observations remain the same. However, the Laplace approximations conducted by INLA require iterative optimization steps to find modes of high-dimensional functions, and in some cases these optimization steps may be substantially more computer-intensive for a simpler model, for instance when the mode is relatively hard to identify. Therefore, there is no contradiction in the reported results.

\begin{table}[ht]
\centering
\resizebox{\textwidth}{!}{ \begin{tabular}{|c|c|c c c|c|} 
 \hline
 Model number & Model form & WAIC & CPO & RMSE$_{\text{CV}}$ & Run-time\\ 
 \hline
 1  & $x\cdot\alpha(s-s_0,t-t_0) + \{Z^0(s,t)\}$ & 108 &  -0.0003 & 0.001 &  99\\  
 3 & $x\cdot\alpha(s-s_0,t-t_0) + \gamma(s-s_0,t-t_0) + \{Z^0(s,t)\}$ & 59 & 0 & 0 & 206\\  
 4 & $x\cdot\alpha(s-s_0,t-t_0) + \gamma(s-s_0,t-t_0) + x^\beta \cdot \{Z^0(s,t)\}$ &  0  & -0.0018 & 0.091 & 71\\ 
 5 & $x\cdot\alpha(s-s_0,t-t_0) + x^\beta \cdot \{Z^0(s,t)\}$ &  47 & -0.0004 & 0.094 & 89\\
 \hline
\end{tabular}}
 \caption{Summary of conditional space-time models, model selection criteria, and total run-times (minutes). The minimum WAIC value ($-215973$ for Model~4); maximum CPO value ($2.92$ for Model~3); and minimum RMSE$_\text{CV}$ value ($1.09$ for Model~3) have been subtracted from their respective columns. We estimate $\beta$ as $0.55$ with a 95\% credible interval of $(0.49,0.64)$ (Model~4) and $0.55$ $(0.50,0.65)$ (Model~5).}
 \label{tab:spacetimeResults}
\end{table}

\section{Computational and implementation details}\label{sec:computation}

\subsection{Introduction}
This section provides further details on INLA, the SPDE approach, and specifics of implementation that are necessary to gain a full understanding of our methods, but not to appreciate the general ideas behind the approach.

\subsection{Bayesian inference with the integrated nested Laplace approximation}
    The integrated nested Laplace approximation \citep[INLA;][]{Rue.al.2009,Rue.al.2017,Opitz.2017b,Niekerk.et.al.2019} provides relatively fast and accurate analytical approximations for posterior inference in models with latent Gaussian processes. The distribution of the observed variables may be non-Gaussian conditional on the latent Gaussian process. \edit{Although here the focus of our modeling approach for conditional extremes is on Gaussian responses,} this does not imply a joint Gaussian assumption on our data, as explained in Section~\ref{sec:spatialModels}. The method astutely combines  Laplace approximations \citep{Tierney.Kadane.1986}, used  to compute expectations with respect to high-dimensional multivariate Gauss--Markov random vectors (denoted by $\bm W$ in Section~\ref{sec:generalities}, with up to tens of thousands of components), with efficient numerical integration schemes for integration with respect to a relatively small number of hyperparameters (denoted by $\bm \theta$) governing variance and \edit{correlation} of Gaussian components, and the shape of the distribution of observations. Therefore, it bypasses issues that may arise with simulation-based Markov chain Monte Carlo (MCMC) inference, where the design of stable algorithms for fast exploration of the posterior distribution may be hampered by intricate dependencies between the components of the model \citep[e.g.,][]{Rue.Held.2005}. With Gaussian distributions for the likelihood as in our model assumption in \eqref{eq:modelgeneral}, the Laplace approximation is exact. INLA is implemented in the \texttt{INLA} package \citep{Lindgren.Rue.2015} of the \texttt{R} statistical software, also referred to as \texttt{R-INLA}, and over the last decade it has been widely adopted for Bayesian additive regression modeling of spatial and spatio-temporal data due to its integration with the stochastic partial differential equation (SPDE) approach \citep{Lindgren.al.2011,Krainski.al.2018}, which provides convenient Gauss--Markov approximations to the Mat\'ern covariance function.  The Bayesian framework of INLA allows for joint estimation and uncertainty assessment of latent components, hyperparameters and predictions. Recently, the speed and stability of INLA with high-dimensional latent Gaussian structures were further leveraged through its integration with the sparse matrix computation library \texttt{PARDISO} \citep{Niekerk.et.al.2019}. 
    \edit{We further point out that the approach of generalized additive models with quadratic penalty terms on coefficients $\bm W$ using frequentist instead of Bayesian estimation of hyperparameters $\bm \theta$  puts no prior distribution on $\bm \theta$,  but the interpretation of $\bm W$ as a Gaussian process is maintained to provide joint estimation of hyperparameters $\bm\theta$ and regression coefficients $\bm W$ through Laplace approximation \citep{Wood2011}, similar to the INLA method.} 

\edit{For a concrete example of how Laplace approximation of integrals  representing posterior estimates (i.e., certain expectations) works, we show how to use it for obtaining the posterior distribution of $\theta_1$, the first component of the hyperparameter vector $\bm\theta$. We denote by $\bm{\theta}_{-1}=(\theta_2,\ldots)^T$ the hyperparameter vector with the component to estimate removed. Since the function arguments are considered as non-stochastic, we here use lower-case notation $\bm v$ and $\bm w$ for $\bm V$ and $\bm W$, respectively.  We have 
$$
\pi(\theta_1\mid \bm v) = \int_{\bm \Theta_{-1}}\int_{\mathbb{R}^m} \pi(\bm \theta,\bm w \mid \bm v) \,\mathrm{d}\bm w\mathrm{d}\bm{\theta}_{-1},
$$
where the outer integral $\mathrm{d}\bm{\theta}_{-1}$ can be disregarded if $\bm{\theta}=\theta_1$ has only one component. The joint posterior density of $\bm w$ and $\bm \theta$ can be calculated up to a  constant as follows, 
$$
\pi(\bm \theta,\bm w \mid \bm v)  \propto \pi(\bm v \mid \bm \theta, \bm w) \times \pi(\bm w\mid \bm \theta)\times \pi(\bm\theta) = \exp\left(\log \pi(\bm w\mid \bm \theta) + \sum_{j=1}^d \log \pi(V_j \mid \bm \theta, \bm w)\right) \times \pi(\bm\theta), 
$$
where the proportionality factor $\pi(\bm v)^{-1}$ is constant for a fixed dataset $\bm v$. 
Writing 
$$
g(\bm w)= \log \pi(\bm w\mid \bm \theta) + \sum_{j=1}^d \log \pi(v_j \mid \bm \theta, \bm w)
$$ 
for the function in the exponent, we replace it by a quadratic approximation using its second-order Taylor expansion around its modal configuration $\bm w^\star$ with $g(\bm w^\star) = \max_{\bm w} g(\bm w)$, i.e., 
$$
g(\bm w) = g(\bm w^\star) + (\bm w-\bm w^\star)^T g''(\bm w^\star) (\bm w-\bm w^\star)
$$
with the Hessian matrix $g''(\bm w^\star)$ of $g$. This defines an approximation of the integrand $\pi(\bm \theta,\bm w \mid \bm v)$ via a multivariate Gaussian density, and therefore the value of the integral with respect to $\mathrm{d}\bm w$ can be calculated straightforwardly. If the likelihood $\pi(v_j\mid \bm \theta, \bm w)$ is Gaussian, then the approximation of $g''(\bm w^\star)$ through a Gaussian density is exact and easy to calculate directly.  In the general case, numerical implementations such as  the \texttt{R-INLA} software use iterative algorithms to find $\bm w^\star$. Finally, the outer integral with respect to  $\mathrm{d} \bm \theta_{-1}$ in relatively small dimension is calculated through an appropriate discretization scheme. Note that the Laplace approximation of the inner integral has to be calculated for each discretization point of $\bm \theta_{-1}$. A similar approximation scheme can then be applied for posterior densities of some component $w_j$ of $\bm w$, or of some linear combination of components of $\bm w$ (e.g., components of the linear predictor $\bm \eta$), where Laplace approximation is used to calculate the integral with respect to $\mathrm{d}\bm w_{-j }$. When the likelihood is Gaussian as in our case,  then one can simply use the exact conditional distributions $\pi(w_j \mid \bm w_{-j}, \bm v, \bm \theta)$, which are univariate Gaussian.  }

\subsection{The SPDE approach}\label{subsec:SPDE}
The latent variable framework allows us to choose the spatial resolution of the latent model separately from that of the observed locations. Moreover, we can use the results of \citet{Lindgren.al.2011}, known as the stochastic partial differential equation (SPDE) approach, to work with numerically convenient Markovian approximations to the Mat\'ern covariance function, leading to sparse precision matrices. We consider random fields defined on $\mathbb{R}^D$; for the residual process $\{Z^0(s)\}$, $D=2$, but we will also use this framework with $D=1$ to define the spline functions with respect to the distance to the conditioning site. The SPDE is given by
\begin{equation}
\label{eq:spde}
\left(\kappa^2 - \Delta \right)^{\zeta/2} \tau \{W(s):s\in \mathbb{R}^D\}  = \{B(s):s\in \mathbb{R}^D\},
\quad  \zeta=\nu+D/2, 
\end{equation}
with the Laplace operator $\Delta y=\sum_{j=1}^D \partial^2 y/\partial^2 x_j$,  a standard Gaussian white noise process $\{B(s)\}$, and parameters $\kappa>0$ (controlling correlation range) and $\tau>0$ (controlling the variance). It has a unique stationary solution given by a zero-mean Gaussian process $\{W(s)\}$ with Mat\'ern covariance function. Here, $\nu$ is the shape
parameter of the  Mat\'ern, with
$\nu=0.5$ yielding the exponential covariance model.  The marginal
variance \edit{of $\{W(s)\}$ is $\sigma_{Z}^2=\Gamma(\nu)/(\Gamma(\nu+D/2)(4\pi)^{D/2}\kappa^{2\nu}\tau^2)$,
and the \emph{empirical range}, where a correlation of approximately $0.1$ is attained between two
points, is approximately $\rho_{Z}=\sqrt{8\nu}/\kappa^2$}. Note that this range parameter is different from the range in the classical Mat\'ern parametrization. 

In practice, the domain is finite, i.e., different from $\mathbb{R}^D$, and appropriate boundary conditions must be imposed to ensure a solution that is unique in terms of finite-dimensional distributions.  An approximation to the exact solution satisfying the boundary conditions is constructed through the representation $W(s) = \sum_{j=1}^{m} W_j \Psi_j(s)$ with locally supported basis functions $\Psi_j(s)$ (e.g., linear or quadratic B-splines for $D=1$, and finite elements for $D=2$). The basis functions do not depend on SPDE parameters. The stochastic solution $\{W(s)\}$ of the SPDE in the subspace of functions spanned by the linear combination of basis functions then yields $\bm{W} = (W_1,\ldots,W_{m})^T\sim \mathcal{N}_{m}(0,Q^{-1})$ with precision matrix $Q$ known in analytical form. \edit{We emphasize that $\{W(s)\}$ here could represent the splines used for $\alpha(s-s_0)$ or $\gamma(s-s_0)$, or the spatial process $\{Z(s)\}$ used in the construction of $\{Z^0(s)\}$.} In Section~\ref{sec:proposed.a}, we labelled the corresponding latent variables $\bm{W}_\alpha\in\mathbb{R}^{m_\alpha}$, $\bm{W}_\gamma\in\mathbb{R}^{m_\gamma}$ and  $\bm{W}_Z\in\mathbb{R}^{m_Z}$. For $D=2$, we use Neumann boundary conditions where the outward derivative of the realizations of the Gaussian field is zero, which is the default choice for spatial modeling with \texttt{INLA}. For $D=1$ and a support given by an interval, a unique approximation to the SPDE solution exists with free boundaries. In our models where spline functions are constrained to value zero at the origin, we use constructions with a Dirichlet boundary on the left side of the interval, such that the solution satisfies the constraint. Theoretical results in \citet{Lindgren.al.2011} show that the approximation to the  solution is good in general and can be made arbitrarily close by choosing a finer finite element mesh.

The value of $\zeta$ in \eqref{eq:spde} determines how the approximate solution of the SPDE can be constructed in practice \citep{Lindgren.al.2011}, and it must be fixed when estimating the model with INLA. The INLA implementation currently supports using $\zeta\in[1,2]$, i.e., $\nu\in[0,1]$ for $D=2$. 

The vector $\bm{W}_Z$ contains the variables used to represent a single replicate of the Gaussian process. When modeling conditional extremes,  we usually extract $n>1$ extreme episodes satisfying $X(s_0)>u$. To represent the unconstrained residual spatial process $\{Z(s)\}$, we therefore need  independent replicates $\bm{W}_{Z,j}$, $j=1,\ldots,n$, of $\bm{W}_Z$. Moreover, for the purpose of space-time modeling, we may assume that  single episodes span $\ell\geq 1$ time steps. Then, for the unconstrained residual process $\{Z(s)\}$ associated with each episode, we will define a Gaussian vector with $\ell\times m_Z$ components, and there will be $n$ replicates of this vector. We will write the precision matrices of Gaussian vectors comprising several blocks of the initial variables $\bm{W}_Z$ through Kronecker products of matrices; see Section~\ref{subsec:spacetime}.  

\subsection{Imposing the condition $Z^0(s_0)=0$ on the residual process}\label{subsec:residualConstraint}
As mentioned in Section~\ref{subsec:CSEintro}, the residual process $\{Z^0(s)\}$ in the spatial conditional extremes model can be constructed by starting with a Gaussian process $\{Z(s)\}$ and imposing the ($Z(s_0)=0$)-constraint in some way. \cite{Wadsworth.Tawn.2019} propose two options: either subtract the value at the conditioning site, i.e., set the residual process to be $\{Z(s)\}-Z(s_0)$; or use the conditional process $\{Z(s)\}\mid Z(s_0)=0$.

In the latent variable framework, we can obtain a residual process of form $\{Z(s)\}-Z(s_0)$ without losing the latent Markovian structure, since we only need to manipulate the representation for $\{Z(s)\}$, which has a sparse precision matrix. The latent variables representing $\{Z(s)\}$ are handled as usual, but we modify the observation matrix $A_S$ of the spatial process $\{Z(s)\}$ to obtain $A_S^0$, the observation matrix associated with the process $\{Z(s)\}-Z(s_0)$. Therefore, let $A_{s_0}$ denote the observation matrix for the conditioning site of dimension $1\times m_Z$, and $A_S$ the observation matrix for the observation locations with dimension $d\times m_Z$. Then, we apply the transformation
\begin{equation}\label{eq:Aspatial}
A_S^0 = A_S - \begin{pmatrix} A_{s_0} \\ \vdots \\  A_{s_0} \end{pmatrix} \in \mathbb{R}^{d \times m_{Z}}
\end{equation}
to obtain the new observation matrix. 

The alternative approach is to impose the ($Z(s_0)=0$)-constraint via conditioning, in the sense of the conditional probability distribution. In general, if $\bm W \sim \mathcal{N}_m(\bm 0,Q^{-1})$ is an $m$-dimensional Gaussian random vector with precision matrix $Q$, we may want to impose a linear constraint of the form
$$
B \bm W = \bm e, \quad B\in\mathbb{R}^{k\times m}, \quad \bm e \in \mathbb{R}^k, 
$$
where $k$ is small.
For instance, $B=(1,0,\ldots,0)$ and $\bm e=0$ if we constrain the Gaussian vector to satisfy $W_1=0$, or $B=(1/m,\ldots,1/m)$ and $e=0$ if we constrain the average value to $0$. 
The linear transformation
$$
\bm W\mid (B\bm W = \bm e) \quad \stackrel{d}{=} \quad \bm W - Q^{-1} B^T\left(BQ^{-1}B^T\right)^{-1}(B\bm W -\bm e)
$$
of the unconstrained vector $\bm W$ imposes this constraint in the sense of generating a realization of the conditional distribution given $B\bm W = \bm e$. In practice, one can calculate $BQ^{-1}$ by solving $k$ linear systems without explicitly calculating and storing $Q^{-1}$, and fast implementations exist when $Q$ is sparse and $k$ is very small. This approach is known as \emph{conditioning by kriging} \citep[see, e.g., equation (8) in ][]{Rue.al.2009,Cressie.1993}; it is available in \texttt{R-INLA}, and we use it for the implementation of the models presented here. Another possibility, applicable in a more specific setting by allowing us to directly condition the Gaussian vector $\bm W$ on $W_1=0$ (here using the first component without loss of generality), is to remove $W_1$ from $\bm W$, resulting in $\bm W_{-1}$. The precision matrix of $\bm W_{-1}$ conditional on $W_1=0$ then corresponds to $Q$ but with the first row and the first column removed. Since this approach is less general (specifically, in order to impose $Z^0(s_0)=0$, we require that a knot is placed at $s_0$), we here prefer the approach of conditioning through kriging. With respect to model structure, the difference between the two approaches is that conditioning through kriging does not fix the constraint in the prior model, but imposes it in the posterior model by applying the conditioning transformation during the Laplace approximations of INLA, while the second approach directly fixes the constraint in the prior model. In both cases, the condition is appropriately incorporated into the posterior model, and no notable differences arise in the posterior models returned by \texttt{R-INLA}.

For our Red Sea data application, we found that the choice of residual process does not have a large impact on results. The option of using the form $\{Z(s)\}-Z(s_0)$ performed slightly better overall, and we therefore used this method for the results presented in Sections~\ref{sec:spatialInference} and~\ref{sec:spacetimeInference}. A comparison of results using the two different approaches is provided in Appendix~\ref{app:Z0comparison}.

\subsection{Space-time Gauss-Markov models}\label{subsec:spacetime}

Inference on spatial conditional extremes is usually based on replicated observations, corresponding to extreme events of the spatial process $\{X(s)\}$, and in the case of space-time conditional extremes on replicated observations of extreme episodes stretching over several time steps. In this section, we detail how to combine Kronecker products of precision matrices, appropriate observation matrices, and the conditioning approaches outlined in Section~\ref{subsec:residualConstraint}, to generate the latent variable representations of the residual processes $\{Z^0(s)\}$ using sparse precision matrices.

In a setting with  $\ell \geq 1$ independent and identically distributed replicates of spatial Gaussian fields, the joint precision matrix of the $\ell$ fields considered at a fixed set of spatial locations can be represented as the Kronecker product $Q_{ST} = I_\ell \otimes Q_S$, where $I_\ell$ is the $\ell\times \ell$ identity matrix and $Q_S$ is a purely spatial precision matrix. More general time-stationary but temporally dependent sparse precision matrices are possible using the assumption of separable space-time dependence. Given sparse precision  matrices $Q_S$ and $Q_T$, the latter representing the purely temporal covariance structure, the precision matrix for $\ell$ time steps of the space-time process corresponds to the Kronecker product $Q_{ST} = Q_T \otimes Q_S$. The precision matrix  $Q_T$ corresponds to a stationary Gaussian time series (e.g., of a first-order auto-regressive process), assumed to have variance $1$ for the sake of identifiability of variance parameters; see Section~\ref{sec:implementation} for further details.

With \texttt{R-INLA}, the standard choice for modeling spatio-temporal dependence is temporal auto-correlation for $Q_T$. Using discrete and equidistant time steps, we consider the stationary space-time process $\{W(s,t)\}$, with auto-correlation parameter $\rho\in(-1,1)$, given as
\begin{align}
\{W(s,1)\} &= \{\varepsilon_1(s)\}, \notag \\
\{W(s,t+1)\} &= \rho \{W(s,t)\} + \sqrt{1-\rho^2}\{\varepsilon_{t+1}(s)\}, \quad t=1,2,\ldots, \label{eq:ar1}
\end{align}
where $\varepsilon_t$, $t=1,2,\ldots$ are Gaussian random fields with Mat\'ern covariance, and $\{W(s,t)\}$ and $\{\varepsilon(s,t)\}$ possess the same variance. \edit{In our setting, this auto-regressive structure is only used to model temporal dependence within single extreme episodes, and there is no assumption of dependence between different extreme events.} The space-time precision matrix for the Cartesian product of a collection of sites and times corresponds to the Kronecker product of the corresponding purely spatial Mat\'ern precision matrix $Q_S$, and the purely temporal $\ell\times\ell$ precision matrix $Q_T^{\mathrm{AR1}}$ of a stationary first-order auto-regressive process with marginal variance $1$, defined as follows for $\ell\geq 1$ time steps:
$$
Q_T^{\mathrm{AR1}} = \frac{1}{1-\rho^2}
\begin{pmatrix}
1 & -\rho & & & & \\
-\rho & 1+\rho^2 & -\rho & & & \\
& -\rho & 1+\rho^2 & -\rho & & \\
& & \ddots & \ddots & \ddots &  \\
& & & -\rho & 1+\rho^2 & -\rho \\
& & & & -\rho & 1
\end{pmatrix}.
$$
The Kronecker product $Q_T^{\mathrm{AR1}}\otimes Q_S$ then has the following form:
$$
Q_{ST} = \frac{1}{1-\rho^2}
\begin{pmatrix}
Q_S & -\rho Q_S & \cdots & \cdots & 0 \\ 
-\rho Q_S &  (1+\rho^2)Q_S &  -\rho Q_S & \cdots & 0 \\ 
\vdots &  \ddots  &  \ddots & \ddots & \vdots \\ 
0 &  \cdots  & -\rho Q_S &  (1+\rho^2)Q_S& -\rho Q_S \\ 
0 &  \cdots  &   \cdots  &  -\rho Q_S & Q_S 
\end{pmatrix}.
$$
We can modify the spatio-temporal Gaussian process to enforce  $Z^0(s_0,t_0)=0$ in the corresponding residual process by analogy with the spatial setting in Section~\ref{subsec:residualConstraint}. The procedure for conditioning on $Z(s_0,t_0)=0$ also does not present notable differences, and we now detail the alternative approach of using the construction $\{Z^0(s,t)\}=\{Z(s,t)\}-Z(s_0,t_0)$. Assume, without loss of generality, that the time $t_0$ with the observed conditioning value corresponds to the first time step of each extreme episode, as outlined in Section~\ref{subsec:RedSea_spacetime}, and that the same locations are observed during the $\ell$ time steps. We first define $A_{s_0,t_0}$ as the observation matrix for the conditioning site and time with dimension $1\times (m_Z\times \ell)$, where $m_Z$ is the number of latent variables for a single spatial replicate, as before. For instance, if $Z(s_0,t_0)$ corresponds to the first latent variable, then $A_{s_0,t_0}=(1,0,\ldots,0)$.
Using the notation for spatial observation matrices as defined in \eqref{eq:Aspatial}, the observation matrix $A_{ST}^0$ for one episode of the residual process $\{Z^0(s,t)\}=\{Z(s,t)\}-Z(s_0,t_0)$ 
is given by the modified block-diagonal matrix 
$$
A_{ST}^0 = 
\begin{pmatrix}
A_S & 0 & \cdots & \cdots & 0 \\ 
0 &  A_S &  0& \cdots & 0 \\ 
\vdots &  \ddots  &  \ddots & \ddots & \vdots \\ 
0 &  \cdots  & 0 &  A_S& 0 \\ 
0 &  \cdots  &   \cdots  &  0 & A_S 
\end{pmatrix}
-
\begin{pmatrix}A_{s_0,t_0} \\ \vdots\\ \vdots \\ \vdots \\ A_{s_0,t_0}\end{pmatrix},
$$
 with $\ell$ blocks on the diagonal, and one or several columns with the same non-negative entries for all rows to represent the term $-Z(s_0,t_0)$. Then, the representation $A_{ST}^0$ coincides with $A_S^0$ in the case of purely spatial extreme episodes ($\ell=1$).

Finally, we take into account the replication structure with $n$ observed replicates of extreme spatial or spatio-temporal episodes.
By assuming that each replicate has  the same design of spatial locations observed over $\ell$  time steps, we can write the overall observation matrix as the Kronecker product $A_{\mathrm{repl}}=I_n\otimes A_{ST}^0$ with the $n\times n$ identity matrix $I_n$.

We emphasize that we use the constructions of spatio-temporal processes based on  \eqref{eq:ar1} for two purposes. First, we can specify the residual process $\{Z^0(s,t)\}$ by using $\{Z^0(s)\}$, with $s\in\mathcal{S}\subset\mathbb{R}^2$, in~\eqref{eq:ar1}. \edit{Second, we can define the  prior structure for the functions $\alpha(s-s_0,t-t_0)$ or $\gamma(s-s_0,t-t_0)$ by using independent copies of  $\alpha(s-s_0)$ or $\gamma(s-s_0)$, respectively, for the innovation process $\varepsilon$ in \eqref{eq:ar1}, with $t$ running from $1$ to  $\ell$.} The latter case can be seen as the use of a Gaussian process prior for the coefficients of a tensor product spline basis, defined with respect to the dimensions of spatial distance and time lag. 

The form of the process $\{W(s,t)\}$ used here exhibits separable dependence in space and time. At present, there are no other, more flexible non-separable models indexed over continuous space and readily implemented within \texttt{R-INLA}, although the possibility of such models has been discussed by \citet{Bakka.al.2018}. \edit{\cite{Simpson.Wadsworth.2020} consider the case for using non-separable dependence forms within spatio-temporal conditional extremes models. They conclude that allowing for non-separability in the normalizing functions (e.g., where $a_{(s,t)}$ cannot be decomposed additively or multiplicatively into a purely spatial and a purely temporal term) is more important than in the residual process} since these capture more of the structure in the model. Within \texttt{R-INLA}, the semiparametric specification of the first normalizing function $a_{(s,t)-(s_0,t_0)}$ allows for flexible, non-separable structure in the posterior estimate of the function (which is not to be confounded with the fact that its prior dependence structure is separable). More complex, non-separable parametric forms of the function $b_{(s,t)-(s_0,t_0)}$ could be estimated by analogy with the spatial case. We conclude that using a separable form of $\{W(s,t)\}$ here should be sufficient. 

\subsection{Implementation using the \texttt{R-INLA} software}\label{sec:implementation}

\edit{For examples of the implementation of our proposed approach in \texttt{R-INLA}, we have made annotated code available on GitHub (\url{https://github.com/essimpson/INLA-conditional-extremes}). When fitting models using this approach, the default \texttt{R-INLA} output consists of discretized versions of univariate posterior densities for the components of the latent Gaussian model terms (including the linear predictor components)  and hyperparameters of the model. Moreover, standard summaries of these univariate posteriors are provided, including posterior means, medians and $95\%$ credible intervals. It is also possible to request further specific outputs when calling the \texttt{inla} function, which carries out the estimation in \texttt{R}, such as estimates of the WAIC and CPO values. If required, one can also obtain approximate posterior samples after having estimated the model, which allows for posterior inference using Monte Carlo estimation for more complex quantities that are not part of the output that can be provided directly.}   

While standard functionality available in the \texttt{R-INLA} library allows for straightforward implementation of the unconstrained SPDE model and auto-regressive structures for dimensions $D=1,2$, as presented in Sections~\ref{subsec:SPDE} and \ref{subsec:spacetime}, respectively, more specific extensions are required for imposing the condition $Z^0(s_0,t_0)=0$.

The \texttt{R-INLA} package provides the precision matrices of the unconstrained latent spatial process $\{Z(s)\}$. Space-time processes $\{Z(s,t)\}$, and independent replications of spatial or spatio-temporal processes, are then handled internally by \texttt{R-INLA}. To estimate  model components of type \edit{$x^\beta\left[\{Z(s,t)\}-Z(s_0,t_0)\right]$ where the parameter $\beta$ does not need to be estimated through INLA}, we can simply modify the observation matrix $A_{\mathrm{repl}}$ and  give it as an input to the estimation routine. As to imposing the  constraint where we condition on $Z(s_0,t_0)=0$, the conditioning-by-kriging approach using a matrix $B$ and a vector $\bm e$ is already implemented for the spatial $\{Z(s)\}$ process ($D=2$), and can be used for spatial extreme episodes with $\ell=1$. Similarly, for $D=1$ the condition $Z(0)=0$ can be set through a flag in \texttt{R-INLA}, and we will deploy this mechanism to  constrain priors of spline functions used to model the functions $\alpha(s-s_0)$ and $\gamma(s-s_0)$ in \eqref{eq:modelgeneral}. 
However, space-time models ($\ell>1$) with temporal auto-regression, where the condition is active only for exactly one of the $\ell$ time steps, are not possible through this mechanism in \texttt{R-INLA}. Similarly, \edit{variances of the residual space-time process $x^\beta\{Z^0(s,t)\}$  that vary over the $\ell$ time steps, with non-stationarity expressed through hyperparameters to be estimated, are not directly available.}

Many additive components of the latent model that are not directly available through standard mechanisms in \texttt{R-INLA} can be implemented manually through its \texttt{rgeneric} function. This requires us to manually define functions that return the precision matrix, the (deterministic) mean function (if different from $0$), and the prior densities of the hyperparameters of the component to be set up. 
\edit{In particular, the parameter $\beta$ may be treated as a hyperparameter to be estimated.} Moreover, \edit{we could estimate a parametric mean function $a_{(s,t)-(s_0,t_0)}(x)$ in $a_{(s,t)-(s_0,t_0)}(x)+x^\beta\{Z^0(s,t)\}$}, where $a_{(s,t)-(s_0,t_0)}(x)$ depends on hyperparameters but does not involve any of the latent Gaussian components gathered in the vector $\bm W$. Finally, the conditioning on  $Z(s_0,t_0)=0$ in the  spatio-temporal setting ($\ell>1$) can be imposed by combining an unconditional  \texttt{rgeneric}-model with the conditioning through kriging technique available as a standard mechanism within \texttt{R-INLA}. In all operations involving large precision matrices, it is crucial to use appropriate sparse matrix objects and sparse matrix operations in the \texttt{R} language. 

In Section~\ref{subsec:spatialModels}, we proposed a cross validation procedure that involved removing and subsequently predicting observations in a particular region of the spatial domain. We now highlight that in \texttt{R-INLA}, this is straightforward to achieve by replacing the data for the cross validation by missing data flags, since these values will automatically be estimated, e.g., through the posterior mean of the ``fitted values".

\section{Discussion}\label{sec:discussion}
The aim of this paper was to develop an inferential approach for spatial and spatio-temporal conditional extremes models, by exploiting latent Gaussian processes within the SPDE framework, and with efficient inference carried out using \texttt{R-INLA}. A benefit of this method is that we are able to handle more spatial or spatio-temporal locations than is possible using existing likelihood-based techniques. In principle, the Laplace approximations carried out within INLA could also be used for frequentist inference without specifying prior distributions for hyperparameters, but we emphasize that the Bayesian framework comes with some valuable benefits, such as the control of model complexity via the use of penalized complexity priors. High-dimensional inference was facilitated by accepting some modest restrictions on the modeling set up. Firstly, we only considered inference based on a single conditioning location. As mentioned in Section~\ref{sec:sensitivity}, in other contexts sensitivity to the choice of conditioning location has been reduced by use of composite likelihoods to incorporate all potential conditioning locations. However, this comes at a computational cost, with a further much larger cost to assess uncertainty via the bootstrap. Secondly, we only allow for the residual process to have a Gaussian form. In many applications this is likely to be adequate, but may lead to problems if the domain of the data is sufficiently large that there is approximate independence in the extremes at long distances. This is because under independence, we expect $\alpha(s-s_0) =0$, $\gamma(s-s_0) =0$ and $\beta = 0$, such that $\{X(s)\}|[X(s_0)=x] = \{Z^0(s)\}$, but there is a mismatch between the marginals of $\{X(s)\}$ (Laplace) and $\{Z^0(s)\}$ (Gaussian). \citet{Wadsworth.Tawn.2019} dealt with this by allowing more general forms for the margins of $\{Z^0(s)\}$, but where this is not necessary, use of untransformed Gaussian processes is certainly more efficient. In principle, non-Gaussian responses can be handled within \texttt{R-INLA} by using a response distribution (i.e., a ``likelihood model") different from the Gaussian; however, due to the conditional independence assumption with respect to the latent Gaussian process, it may be difficult to obtain models that realistically reflect the spatio-temporal smoothness of observations. In contrast to the existing inferential approach, INLA allows us to estimate flexible semiparametric specifications for the functions arising in the mean of the Gaussian process of conditional extremes, and the estimation and uncertainty assessment is performed jointly with all other model parameters.

Since we construct models using a single conditioning site, $s_0$, but may subsequently assume that the fitted model applies at other conditioning locations, another important consideration is the choice of a suitable position for $s_0$. There are certain aspects to take into account here, as highlighted in Section~\ref{sec:sensitivity}. For instance, one may wish to choose $s_0$ so that $\max_{i=1,\dots,d}\|s_i-s_0\|$ takes its largest value, as this will provide more reliable estimates for the spline functions $\alpha(s-s_0)$ and $\gamma(s-s_0)$ at the longest distances. On the other hand, choosing $s_0$ towards the edge of the spatial domain may mean that it is less likely to be representative of the full set of locations. These two considerations should be balanced in the selection of $s_0$.  Even when inference on parameters has been made using a single conditioning location, our assumption of spatial stationarity means that it is still possible to infer conditional probabilities or expectations for alternative conditioning sites or events. In particular, \citet{Wadsworth.Tawn.2019} demonstrate how to make inference on quantities of the form
\[
\E[g(\{X(s)\})|\max_{1\leq i \leq d} X(s_i)>u],
\]
for a function $g(\cdot)$ of interest, which could be exploited in our setting just as easily.

In other application contexts, the analysis of non-stationarities in conditional extremes with respect to $s_0$ may be of interest, such that the assessment of differences between models fitted at different conditioning locations $s_0$ is an inferential goal in itself. The local modeling suggested by the conditional extremes approach makes sense if we have a large study area with possible non-stationarities, but are mostly interested in inferences on local features. For example, with the Red Sea surface temperature data, one could choose $s_0$ as a representative site of one coral reef, or several closely located coral reefs, although we have not done this here. In climate studies, loosely speaking ``climate" is often considered as pertaining to the characteristics of the marginal distribution, while ``weather" is additionally driven by the local spatio-temporal dependence; with the conditional extremes models, we could also consider the properties of the $s_0$-conditioned model as part of the ``climate" at $s_0$.

As discussed in Section~\ref{sec:sensitivity}, \cite{Wadsworth.Tawn.2019} and \cite{Simpson.Wadsworth.2020} use a composite likelihood approach to combine information across several conditioning sites. While this is also a possibility in our setting, we would lose some of the benefits over the classical likelihood framework since the uncertainty estimation becomes awkward in a Bayesian context, though consistency of estimators is preserved \citep{Soubeyrand2015}. An alternative may be to obtain separate estimates for different conditioning locations, and combine these via some weighted approach, i.e., perform model averaging either in the domain of the models' likelihood or in the domain of their predictions. This provides a potential avenue for further work.

While this was not necessary for our Red Sea data example, it would be relatively straightforward to include covariates within the latent Gaussian structure. For instance, a distant-dependent variance model may be more appropriate in certain cases, and such an adaptation would be possible within the INLA framework by suitable modification of the models outlined in Table~\ref{tab:spatialModels}. As is generally the case with covariate modeling, the difficulty here is in choosing relevant covariates whose influence can be easily interpreted. For some scenarios, the effect of a particular covariate may already be known, and the modeling may benefit from this approach. Another technique that may be useful in certain settings is censoring, which is also possible within the INLA framework via the inclusion of a censored Gaussian response for the likelihood of observations in Section~\ref{sec:generalities}; see  \citet{Zhang.al.2019} for an MCMC-implementation in a similar context.

A further issue linked to modeling non-stationarity is the assumption of isotropy that we place on the underlying spatial process. Indeed, the results in Appendix~\ref{app:chi} show that there is some violation of this assumption in our application. \cite{Wadsworth.Tawn.2019} deal with anisotropy by including a transformation of the spatial coordinates within the modeling procedure. This approach is also adopted by \cite{Simpson.Wadsworth.2020}, and the resulting transformation for the sea surface temperature data in the northern Red Sea is very small. Such a transformation is not incorporated within the standard \texttt{R-INLA} set-up, but the anisotropy parameters could be estimated using a \texttt{generic} model. In the spatial setting, \cite{Richards2021} propose a deformation technique to deal with non-stationarity in extremal dependence features. This allows for anisotropy to be handled as a preliminary modeling step, which it would be possible to do in the context of conditional spatial extremes.

Our approach is not fully Bayesian, since the marginal transformation of data at each spatial location to a standard Laplace distribution is carried out separately to the dependence modeling. This approach appears sufficient, particularly since simultaneous estimation of the margins and the dependence within the INLA framework would be intricate. This would result in us resorting to MCMC estimation, but here we do not want to sacrifice the simplicity and speed of the INLA-implementation with big datasets.

There are several aspects that we have had to consider as part of the implementation of the conditional extremes models in INLA, with some of these being more important than others. We found the priors of the hyperparameters to have minimal importance, as indicated by posteriors with small credible intervals. This is likely due to the large number of observations we had available. We also observed very similar results when setting the SPDE parameter to $\zeta=1.5$ or $\zeta=2$, so for our data, the smoothness of the Gaussian field does not have a significant impact compared to other aspects of the model. As we may have expected, choosing appropriate normalizing functions is hugely important, and the forms of these may need to be tailored to the specific data application.

\paragraph{Data and code} Code to implement the models in this paper is available online in the GitHub repository \url{https://github.com/essimpson/INLA-conditional-extremes}, and the sea surface temperature data can be downloaded at \url{https://marine.copernicus.eu}.


\bibliographystyle{apalike}
\bibliography{biblio}

\appendix
\section{Extremal dependence properties of the Red Sea data}\label{app:chi}
\begin{figure}[!htbp]
\centering
\includegraphics[width=0.75\textwidth]{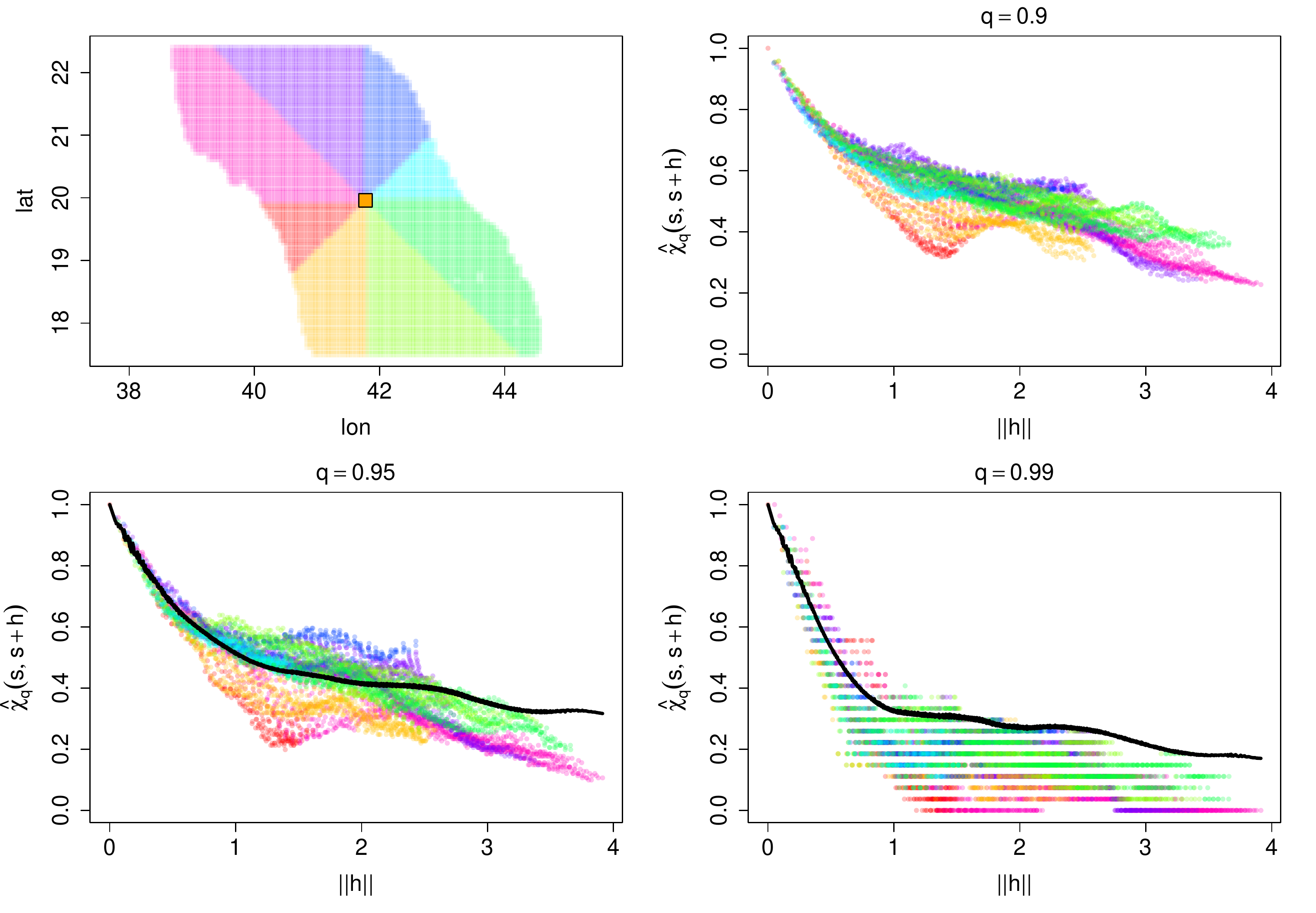}
\label{fig:chis}
\caption{Estimates of $\chi_q(s,s+h)$ for all spatial locations, using one conditioning site (highlighted in orange), and for $q=0.9,0.95,0.99$. The locations in the top-left panel have been divided into eight sections, with the $\chi_q(s,s+h)$ estimates in the remaining plots coloured accordingly. \edit{The black lines in the plots of the bottom row show the corresponding $\chi_q(s,s+h)$ estimates from our fitted Model~3.}}
\end{figure}

In this section, we aim to assess some of the extremal dependence properties of the Red Sea data. To do this, we consider a non-limiting version of the tail correlation function $\chi(s,s+h)$ introduced in equation~\eqref{eq:chi}. In particular, we define
\begin{align*}
\chi_q(s,s+h) = \Pr\left\{F_L(X(s+h))>q \mid F_L(X(s))>q\right\},
\end{align*}
for $F_L$ denoting a standard Laplace distribution, $s,s+h\in\mathcal{S}$ with $X(s), X(s+h)\sim F_L$, and $q\in[0,1]$. We are interested in values of $q$ close to 1 in order to focus on extremal dependence properties. In Figure~\ref{fig:chis}, we provide empirical estimates of $\chi_q(s,s+h)$ for the Red Sea data, with $q=0.9,0.95,0.99$, and $s$ chosen to be the conditioning site $s_0$ used for inference in Sections~\ref{subsec:spatialModels} to \ref{subsec:Model3}. Estimates are shown for $s+h$ being each of the spatial locations in our dataset.

There are two reasons why we are interested in these plots. The first is to assess the type of extremal dependence exhibited by the data, so we can better understand the types of models that will be appropriate here. As the level $q$ increases, it is clear that there is weakening dependence in the data. This suggests that models for asymptotic independence will be more appropriate than those for asymptotic dependence here, a finding that is reflected in the models we favour in both the spatial and spatio-temporal cases. \edit{Estimates of $\chi_{0.95}(s,s+h)$ and $\chi_{0.99}(s,s+h)$ based on simulations from our fitted spatial Model 3 are also included in Figure~\ref{fig:chis}. These demonstrate that the model indeed exhibits dependence that weakens as the threshold increases, albeit at a slower rate than in the empirical estimates.} The second use of these plots is to assess our assumption of isotropy. To do this, the spatial domain have been separated into eight sections, as shown in the top-left panel of Figure~\ref{fig:chis}, and we consider whether the extremal dependence behaviour differs within these sections. It appears that there is some violation of this assumption, since the results in the bottom-left quadrant are different to the others. Some ways to deal with this issue are discussed in Section~\ref{sec:discussion}.

\section{Additional model comparisons}
\subsection{A comparison of the two residual processes}\label{app:Z0comparison}
In Section~\ref{subsec:residualConstraint}, we discussed the implementation of two different structures for the spatial residual process $\{Z^0(s)\}$, both of which ensure the constraint that $Z^0(s_0)=0$ is fulfilled. Throughout the paper, we use a process of the form $\{Z(s)\}-Z(s_0)$, but the alternative is to use $\{Z(s)\}\mid Z(s_0)=0$. In Table~\ref{tab:Z0comparison} we present WAIC results for the models described in Table~\ref{tab:spatialModels} of Section~\ref{subsec:spatialModels} using both residual processes. Models 0 to 5 are applied to the moderate set of spatial locations shown in the right panel of Figure~\ref{fig:spatialLocations} (Model~6 is omitted as it involves no modeling of the residual process), with the threshold $u$ taken as the 0.95 quantile of observations on the Laplace scale. The conditioning site lies towards the centre of the spatial domain, and the mesh and remaining tuning parameters are chosen as previously. Model~3 with the $\{Z(s)\}-Z(s_0)$ residual process, which is the one we focus on in Section~\ref{subsec:Model3}, performs the best here. For any given model, the WAIC values are quite similar using the two residual processes, although the $\{Z(s)\}-Z(s_0)$ construction we have used throughout our analysis is favoured overall. This indicates that the choice of $\{Z^0(s)\}$ is not so critical here, and that we have made a reasonable choice in our modeling.

\begin{table}[ht]
\centering
 \begin{tabular}{|c|cc|} 
 \hline
 Model & $\{Z(s)\}-Z(s_0)$ & $\{Z(s)\}\mid Z(s_0)=0$ \\ 
 \hline
 0 & 337 & \bf 298 \\ 
 1 & \bf 43 & 65 \\ 
 2 & 80 & \bf 71 \\ 
 3 & \bf 0 & 18 \\ 
 4 & \bf 20 & 23 \\ 
 5 & \bf 55 & 61 \\
 \hline
 \end{tabular}
 \caption{Comparison of WAIC values for models fitted with different residual processes. The minimum WAIC value ($-95298$ for Model~3 with residual process $\{Z(s)\}-Z(s_0)$) has been subtracted from all other values. Results in bold demonstrate the minimum WAIC value achieved for each model.}
 \label{tab:Z0comparison}
\end{table}

\subsection{Sensitivity of model fits to SPDE parameter $\zeta$}\label{app:zetaSensitivity}
In Section~\ref{subsec:SPDE}, we gave an introduction to the SPDE approach, with parameter $\nu$ representing the shape parameter of the Mat\'{e}rn covariance. In the SPDE framework, this links to the parameter $\zeta=\nu+D/2$, i.e., in the spatial case $\zeta=\nu+1$. The value of $\zeta$ must be fixed when implementing the INLA methodology, with $\zeta=1.5$ corresponding to an exponential covariance, and $\zeta=2$ providing smoother Gaussian realizations.

In Table~\ref{tab:alphacomparison}, we compare WAIC values for each of the models in Table~\ref{tab:spatialModels}, for $\zeta=1.5$ and $\zeta=2$; all other modeling choices fixed as in Section~\ref{sec:spatialInference}. Model~4 with $\zeta=1.5$ remains the most successful model, although selecting $\zeta=1.5$ or $\zeta=2$ is shown to have little effect for any of the models. We opt to fix $\zeta=1.5$ throughout the paper, as we find this to perform slightly better overall, although the form of the model clearly has more of an effect on the results.

\begin{table}[ht]
\centering
 \begin{tabular}{|c|cc|} 
 \hline
 Model & $\zeta=1.5$ & $\zeta=2$ \\ 
 \hline
 0 & \bf 2438 & 2447 \\ 
 1 & \bf 614  & 624 \\ 
 2 & \bf 743  & 746 \\ 
 3 & \bf 4    & 15 \\ 
 4 & \bf 0    & 12 \\ 
 5 & \bf 611  & 639 \\ 
 6 &  4394961 & \bf 4394960 \\ 
 \hline
\end{tabular}
 \caption{Comparison of WAIC values for models fitted with different values of $\zeta$. The minimum WAIC value (-1460982 for Model~4 with $\zeta=1.5$) has been subtracted from all other values. Results in bold demonstrate the minimum WAIC value achieved for each model.}
 \label{tab:alphacomparison}
\end{table}

\section{Additional diagnostic plots}
\subsection{Histogram of PIT values for Model~3}\label{app:pit}
\begin{figure}[!htbp]
    \centering
    \includegraphics[width=0.4\textwidth,trim={0 1.5cm 0 0},clip]{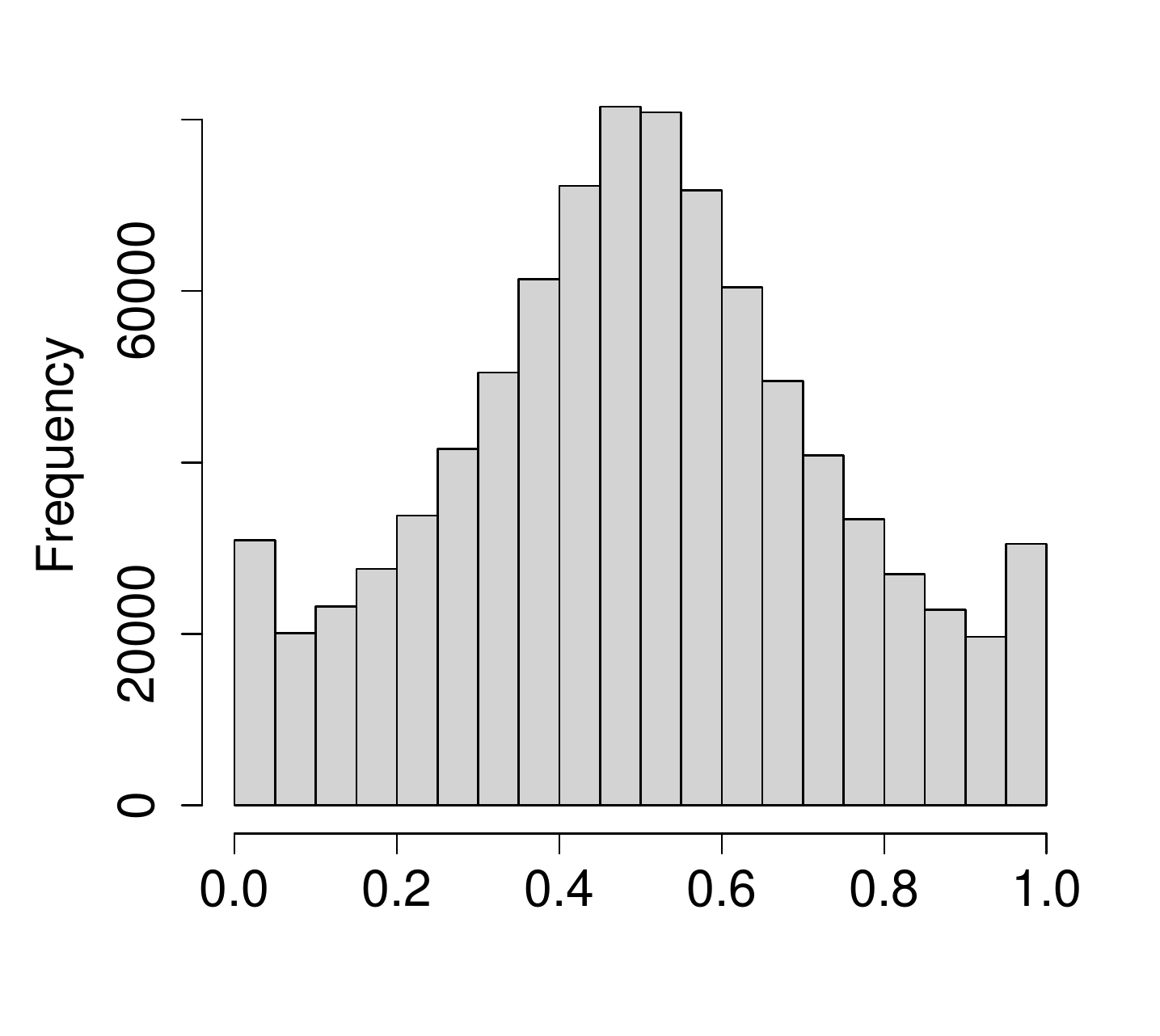}
    \caption{Histogram of PIT values for Model~3.}
    \label{fig:pitHists}
\end{figure}
In Section~\ref{subsec:loocv}, we discussed the use of the probability integral transform as a model fitting diagnostic. The idea is that the closer the PIT values are to being uniform in distribution, the better the model captures variability in predictive distributions when predicting a single data point. In Figure~\ref{fig:pitHists}, we present a histogram of these results for Model~3, with equivalent plots for the Models~0~to~5 being very similar to this one. We recall that since Model~6 does not allow for any residual variation, we are not suggesting it as a serious model contender, and do not consider an equivalent PIT histogram here.

Since we here have smooth, gridded data and a flexible residual process, Models~0~to~5 provide almost perfect predictions \edit{in the setting of leaving out a single observation}. Plots such as this, relating to leave-one-out cross-validation, are therefore less relevant here than in other cases. The peak in the middle of the PIT histogram suggests that the posterior predictive densities $\pi(x_i\mid \bm x_{-i})$ are usually slightly too ``flat" in many cases\edit{,but since the posterior predictive variance is very small for each of the Models 0~to~5, we do not consider this as  problematic. It is therefore useful to consider PIT-based diagnostics alongside information on predictive variance. To wit, this variance is much smaller for Model~3 than Model~6.}

\subsection{\edit{Extension of Figure~\ref{fig:extremeProb1} with extrapolation}}\label{app:extrapolationDiagnostic}
\begin{figure}[!htbp]
    \centering
    \includegraphics[width=0.9\textwidth]{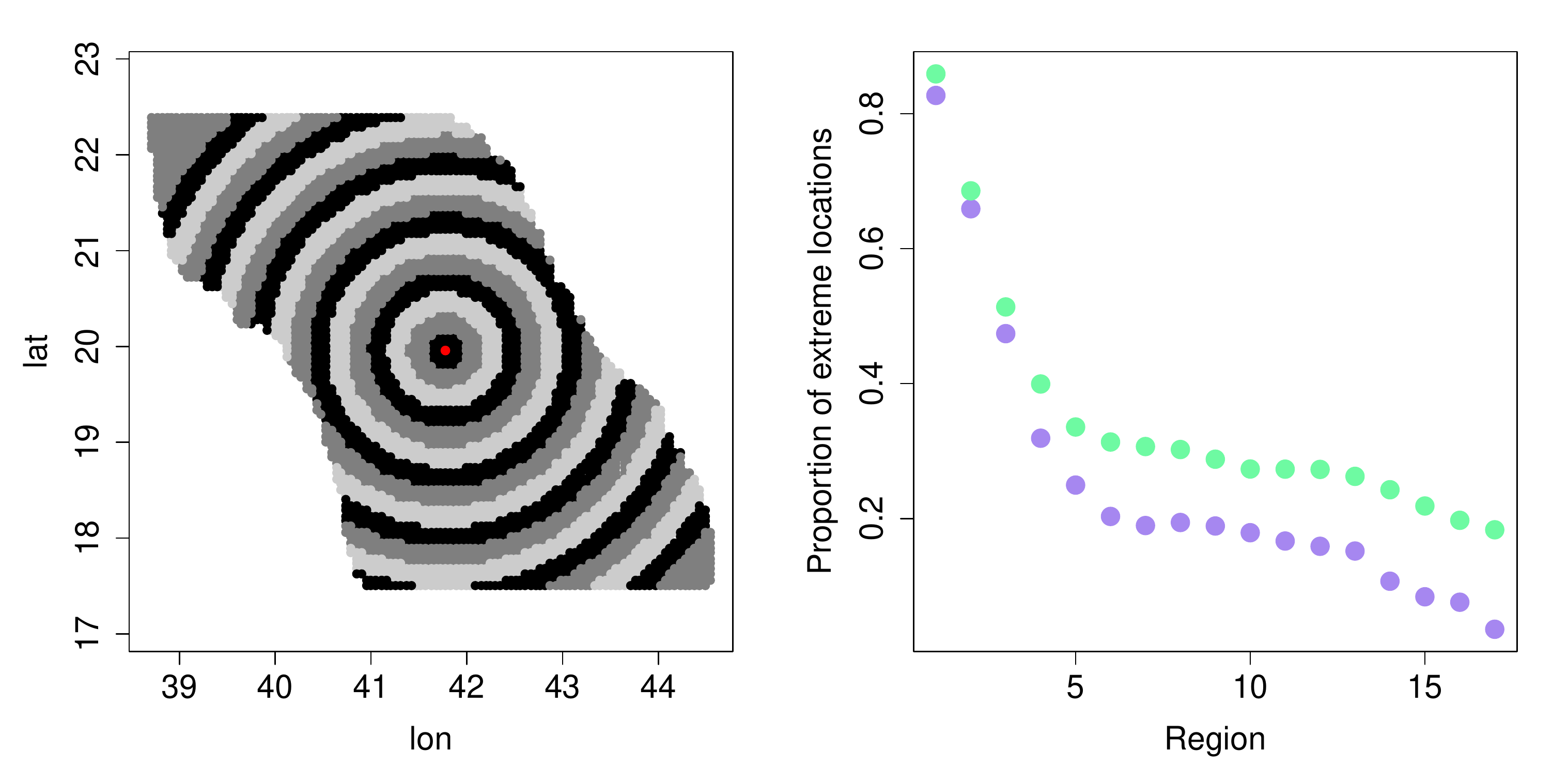}
    \caption{\edit{Left: the spatial domain separated into 17 regions; the region labels begin at 1 in the centre of the domain, and increase with distance from the centre. The conditioning site $s_0$ is shown in red. Right: the estimated proportion of locations that exceed the 0.99 quantile, given it is exceeded at $s_0$ using Model~3 (green) and equivalent empirical results (purple).}}
    \label{fig:extremeProb-extrap}
\end{figure}

\edit{In Figure~\ref{fig:extremeProb1} of Section~\ref{subsec:Model3}, we presented a diagnostic plot demonstrating the proportion of locations that exceed the 0.95 quantile in different regions of the spatial domain, given that it is exceeded at the conditioning site. In Figure~\ref{fig:extremeProb-extrap}, we present an equivalent plot for exceedances of the 0.99 quantile. Model 3 gives reasonable results for the most central regions, but the dependence is overestimated as we move further from the conditioning site. This is discussed further in Section~\ref{subsec:Model3}.}

\end{document}